\documentclass[10pt,aps,prc,superscriptaddress,twoside,twocolumn,nofootinbib,showpacs,preprintnumbers]{revtex4-1}
\usepackage{graphicx,braket,enumitem,amsmath,amssymb,hyperref,mathtools,bbold,bm,etoolbox,siunitx,nicefrac}
\usepackage[normalem]{ulem}
\usepackage[dvipsnames]{xcolor}
\usepackage[utf8]{inputenc}
\robustify
\bfseries
\allowdisplaybreaks
\hypersetup{ 
  pdfnewwindow=true,      
  colorlinks=true,        
  linkcolor=PineGreen,    
  citecolor=PineGreen,    
  filecolor=PineGreen,    
  urlcolor=PineGreen      
}  
\newcommand{\be}{\begin{eqnarray}}
\newcommand{\ee}{\end{eqnarray}}

\begin{document}

\title{Towards the Minimal Spectrum of Excited Baryons}
\author{J. Landay}
\email{jlanday@gwmail.gwu.edu}
\affiliation{Department of Physics, The George Washington University, Washington, DC 20052, USA}
\author{M. Mai}
\email{maximmai@gwu.edu}
\affiliation{Department of Physics, The George Washington University, Washington, DC 20052, USA}
\author{M. D\"oring}
\email{doring@gwu.edu}
\affiliation{
Institute for Nuclear Studies and Department of Physics, 
The George Washington University, Washington, DC 20052, USA}
\affiliation{Theory Center, Thomas Jefferson National Accelerator Facility, Newport News, VA 23606, USA}
\author{H. Haberzettl}
\email{helmut@gwu.edu}
\affiliation{
Institute for Nuclear Studies and Department of Physics, 
The George Washington University, Washington, DC 20052, USA}
\author{K. Nakayama}
\email{nakayama@uga.edu}
\affiliation{Department of Physics and Astronomy, The University of Georgia, Athens, GA 30602, USA}

\preprint{JLAB-THY-18-2818}

\begin{abstract}
In the light baryon sector resonances can be broad and overlapping and are in most cases not directly visible in the cross section data. Automatized model selection techniques that introduce penalties for resonances can be used to determine the minimally needed set of resonances to describe the data. Several possible penalization schemes are compared. As an application we perform a blindfold identification of hyperon resonances in the $\bar{K} N \to K \Xi$ reaction based on the Least Absolute Shrinkage and Selection Operator (LASSO) in combination with the Bayesian Information Criterion (BIC). We find ten resonances --- out of the 21 above-threshold hyperon resonances with spin $J\le 7/2$ listed by the Particle Data Group. In traditional analyses, it is practically impossible to test the relevance of all  resonances and their combinations that may potentially contribute to the reaction. By contrast, the present method proves capable of determining the relevant resonances among a large pool of candidates.
\end{abstract}

\maketitle

\section{Introduction}\label{sec:intro}
One challenge in the phenomenological interpretation of data from scattering or production experiments is the determination of the resonance spectrum. Typically, the quark model predicts more states than are found in experiments, a phenomenon referred to as the {\it missing resonance problem}~\cite{Ronniger:2011td, Ferretti:2011zz, Glozman:1995fu, Bijker:1994yr, Capstick:1986bm}. Pioneering lattice QCD calculations~\cite{Edwards:2012fx, Engel:2013ig, Engel:2010my} obtain the same SU(6)$\times$O(3) symmetry  pattern of the spectrum~\cite{Edwards:2012fx} as in many quark models although the lattice QCD calculations are still carried out at rather large pion masses and without full control of finite-volume effects. In the framework of Dyson-Schwinger and Faddeev equations, several resonances and their properties can be identified with their physical
counterparts~\cite{Chen:2017pse, Eichmann:2016hgl, Lu:2017cln, Eichmann:2016yit}. Chiral unitary approaches operate directly with the physical degrees of freedom --- mesons and baryons --- and explain the masses and properties of some states~\cite{Sadasivan:2018jig, Bruns:2010sv, MartinezTorres:2009cw, Doring:2007rz, Doring:2005bx} although it is clear that not all excited baryons 
can be hadronic molecules. Some bump structures might even be  kinematic effects due to triangle singularities~\cite{Pilloni:2016obd, Samart:2017scf, Debastiani:2017dlz}.

Even if a unique determination of the amplitude were possible --- referred to as a {\it complete experiment}~\cite{Nys:2015kqa, Wunderlich:2014xya, Workman:2010xc, Sandorfi:2010uv} --- the decomposition into partial waves, or multipoles in case of photon-induced reactions, usually requires a truncation~\cite{Wunderlich:2017dby, Workman:2016irf}. Even then, the multipoles are not guaranteed to clearly reveal resonances, especially if obtained from data with statistical and systematic uncertainties because broad, potentially overlapping resonances are difficult to distinguish from the background. 

In principle, one has to test an arbitrary number of resonances and their combinations in the parametrization of partial-wave amplitudes. The goal is to keep the overall number of needed resonances as small as possible, i.e., to find the simplest description of the data within given uncertainties. The number of possible combinations is usually far beyond what can be tested in the traditional way such as by inserting resonances by hand as $s$-channel singularities in $K$-matrix or dynamical coupled-channel approaches, or as Breit-Wigner terms in simpler parametrizations. 

Several techniques have been developed to address this problem. 
In the SAID analysis~\cite{Workman:2012jf, Workman:2012hx, Workman:2012jf, Workman:2012hx, Workman:2016irf, Tiator:2016btt} only the $\Delta(1232)3/2^+$ resonance is explicitly included in the amplitude. If required by data, other resonances can arise through the non-linearity of the unitary coupled-channel amplitude without manual intervention. Notably, in the 2006 SAID solution (SP06) the number of resonances could be significantly reduced without spoiling the description of elastic $\pi N$ scattering~\cite{Arndt:2006bf}.	

Another technique to search for resonances are mass scans~\cite{Arndt:2003ga, Azimov:2003bb, Anisovich:2015gia}. In a given parametrization, additional Breit-Wigner terms are introduced. Their mass is varied in steps and all other parameters are fitted. If this leads to a significant minimum of the $\chi^2$ at a given mass --- preferably for different final states~\cite{Anisovich:2015gia} --- it could possibly be interpreted as a sign for a new resonance. 

The question arises if there are ways of assessing the significance of new resonances. One criterion is given by the $F$-test~\cite{Wilkinson1981} which tests for the significance of new fit parameters, like the Breit-Wigner coupling, or bare coupling of an $s$-channel resonance state in a $K$-matrix or dynamical coupled-channel approach. This method has two practical drawbacks: On the one hand, data from different experiments tend to have systematic inconsistencies so that the resulting fits are never good in the statistical sense (e.g., passing a $\chi^2$ test). The $F$-test will then admit far too many false states. On the other hand, the $F$-test does not relieve one from testing ``by hand'' each new state, or, more precisely, each combination of an unknown number of new states and established ones. 

There are various ``blindfolded'' ways to test new resonances, without the need of manual intervention, that are robust in the sense that they allow for relative model comparison even if the fit cannot be satisfactory in the frequentist's sense. 
Bayesian inference to determine the resonance spectrum was introduced into baryon spectroscopy by the Ghent group~\cite{DeCruz:2012bv,DeCruz:2011xi}. In a related context, the necessary precision of data to discriminate models was determined in Ref.~\cite{Nys:2016uel}. 

Another method for the partial-wave analysis of mesonic systems was presented in Ref.~\cite{Guegan:2015mea}, see also Ref.~\cite{Williams:2017gwf}. The Least Absolute Shrinkage and Selection Operator (LASSO) allows one to generate a series of models with varying partial-wave content. The best model can be selected by additional criteria like cross validation or various criteria from information theory~\cite{Tibshirani2011, hastie2009the, james2013an}. 

In Ref.~\cite{Landay:2016cjw} different models and criteria were compared with LASSO to determine the minimal multipole content for low-energy neutral pion photoproduction. The method was first demonstrated for synthetic data for which the solution was known and then applied to real data. It was found that some $D$-waves are relevant even at lower energies. The cusp parameter was also precisely determined. See also Ref.~\cite{Wunderlich:2016imj} for a related study of dominant partial waves in photoproduction reactions.

Here, we extend the idea further to address the resonance spectrum itself, i.e., we use LASSO to determine the minimal spectrum required to describe a hadronic reaction. Different penalties are tested for synthetic data in which the solution is known. In the second part of the manuscript, we turn to the analysis of real data for the reaction $K^-p\to K\Xi$. This reaction is selected because the database is relatively small but still exhibits problems of data inconsistencies which makes it a suitable candidate for this pilot calculation testing the robustness of LASSO. Also, the resonance content of this reaction was determined ``by hand'' in Ref.~\cite{Jackson:2015dva} and it is particularly illuminating to see how this traditional method compares to the present results.

We expect that the method can be used in hadron spectroscopy in a wide context, e.g., for mesons~\cite{Jackura:2017amb, Molina:2017iaa, Pilloni:2016obd, Mai:2017vot, Kamano:2011ih} or baryons. Light baryon spectroscopy is plagued by wide and overlapping resonances which makes their determination difficult. Groups like Bonn-Gatchina~\cite{Anisovich:2011fc, Collins:2017sgu, Anisovich:2017ygb}, ANL-Osaka~\cite{Kamano:2013iva, Kamano:2014zba, Kamano:2015hxa}, J\"ulich-Bonn~\cite{Ronchen:2015vfa,Ronchen:2014cna, Ronchen:2012eg}, Kent state~\cite{Shrestha:2012ep, Zhang:2013sva, Zhang:2013cua}, DMT and MAID~\cite{Kamalov:2000en, Chiang:2002vq, Tiator:2010rp, Drechsel:1998hk, Drechsel:2007if}, Giessen~\cite{Shklyar:2014kra, Cao:2013psa}, and other groups~\cite{Mart:2017mwj} dedicate much effort to resonance spectroscopy. The reaction considered here, $K^-p\to K\Xi$, is only one of many for strangeness $S=-1$ that have been analyzed by different groups recently~\cite{Fernandez-Ramirez:2015tfa, Kamano:2014zba, Kamano:2015hxa, Zhang:2013sva, Zhang:2013cua}. 

Bayesian priors have been used in the determination of low-energy constants in Chiral Perturbation Theory~\cite{Wesolowski:2015fqa} and to quantify truncation errors~\cite{Furnstahl:2015rha}. Similarly, the LASSO could be useful in selecting relevant low-energy constants in meson-baryon scattering~\cite{Mai:2012dt, Ruic:2011wf, Bruns:2010sv, Mai:2009ce}.

LASSO was also used in Ref.~\cite{Sadasivan:2018jig} in an attempt to actively remove a resonance to explain data conventionally, i.e., with non-resonant background. This turned out to be impossible, favoring the non-conventional, i.e., resonant explanation. In a different context, LASSO has been recently used in the analysis of lattice QCD data via an optical potential~\cite{Agadjanov:2016mao}. In general, LASSO is expected to be particularly relevant in the analysis of lattice QCD calculation because relatively few data points are available for systems with several two-body channels~\cite{Briceno:2017qmb, Guo:2016zep, Wilson:2015dqa, Doring:2013glu, Doring:2011ip} or three-body systems~\cite{Mai:2018djl, Doring:2018xxx, Mai:2017bge,  Briceno:2016mjc, Hammer:2017kms, Hammer:2017uqm}. LASSO could then be used to limit the number of fit parameter and/or relevant two or three-body channels.

This paper is organized as follows. In Secs.~\ref{sec:formalism}, synthetic data, generated from a partial-wave solution with known resonance content, are analyzed with LASSO regularization and using the Bayesian information criterion (BIC). The efficacy of different penalties to recover the resonance spectrum is tested. In Sec.~\ref{sec:real-data}, LASSO is applied to the actual data of the reaction $K^-p\to K\Xi$. 
Our conclusions are presented in Sec.~\ref{sec:Conclusions}. Two Appendices provide expressions for observables used here and figures for the corresponding synthetic data sets.

\section{Analysis of synthetic data}
\label{sec:formalism}
The considered world database for the transition $\bar KN\to K\Xi$ consists of polarized and unpolarized differential cross sections up to a total energy of the system of  ${W\sim 3.0}$~GeV. In this section, however, we work with synthetic data, while in Sec.~\ref{sec:real-data} the actual data are analyzed.

\subsection{Parametrization}
\label{subsec:parametrization}

\begin{figure*}[t]
\begin{center}
\includegraphics[width=0.99\linewidth, trim=0cm 0.5cm 0cm 0cm]{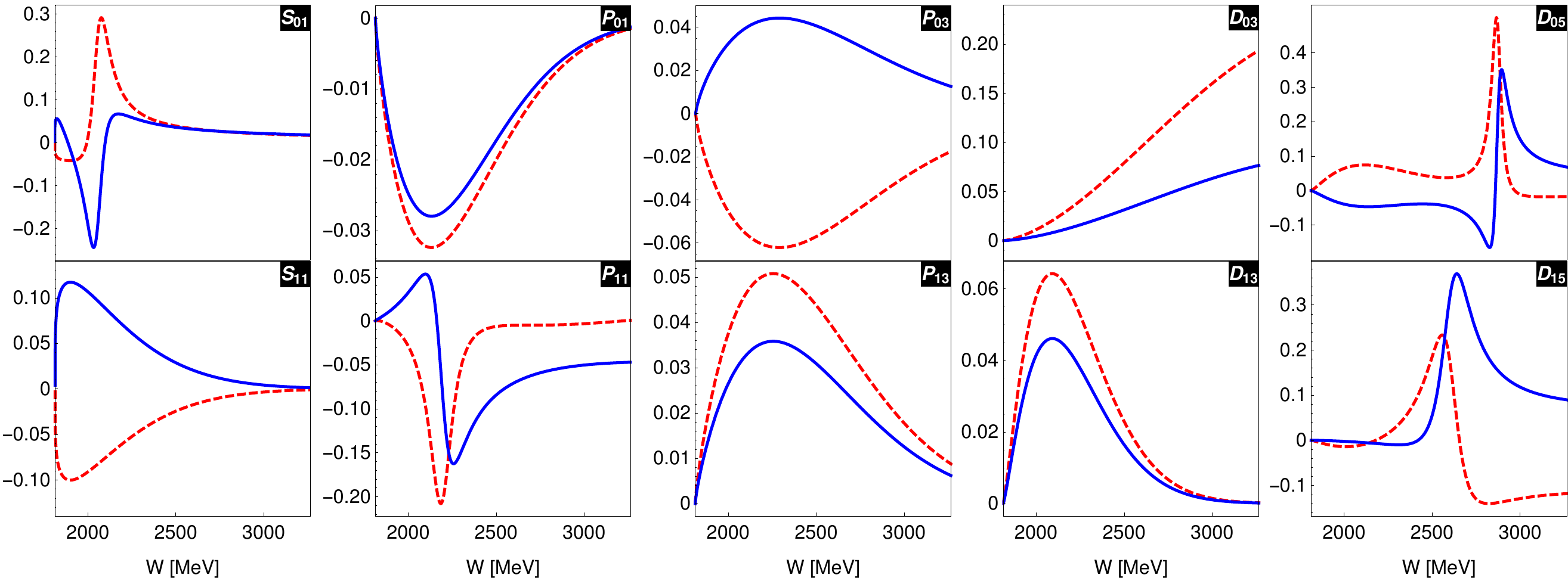}
\end{center}
\caption{%
 Partial-wave amplitudes $\tau$ (in dimensionless units) of the synthetic data for the considered isospin channels. Blue dashed and red solid lines show the real and imaginary parts, respectively.
}
\label{fig:PWA-synthetic}
\end{figure*}

We first consider (synthetic) data addressing the transition $K^-p\to K\Xi$ in terms of the total, polarized and unpolarized differential cross sections. These observables are related to the partial-wave amplitudes (denoted by $\tau$ in the following), as presented in App.~\ref{sec:appA}. For given isospin, total angular momentum, and orbital angular momentum $(I,J,L)$, we assume the following parametrization of the partial-wave amplitudes as a function of energy $W$,
\begin{align}
\label{tau}
\tau(W)&=
e^{i \phi}
 \left(\frac{k_f(W)}{\Lambda} \right)^{L+\nicefrac{1}{2}}\\
&
\mbox{}\quad\times\left(
a\,e^{-\alpha^2 \left(\frac{k_f(W)}{\Lambda}\right)^2} -x\,e^{i\Phi} \frac{\Gamma/2}{W-M+i \Gamma/2} 
\right)\,,
\nonumber
\end{align}
where the scale parameter is fixed as $\Lambda=10^3$~MeV, and $a,\alpha, \phi,\Phi,x, \Gamma,M$ are free (real) parameters for each set of quantum numbers. The final c.m. meson momentum is denoted as $k_f$. We refer to this parametrization as the \emph{benchmark model}. 
In this approach, resonances are introduced as poles with complex residues $x\,e^{i\Phi}$. The correct threshold behavior $\sim k_f^{L+\nicefrac{1}{2}}$ is respected. Also, we cannot exclude relative phases $\phi$ between different partial waves because the amplitude is not real at threshold due to many open channels. Yet, to avoid an overall-phase problem that would make the fit problem ill-defined, we set $\phi=0$ for one partial wave, $(I,J,L)=(1,\nicefrac{5}{2},2)$. 

It should be noted that the present partial-wave parametrization is minimalistic. Properties from $S$-matrix theory like left-hand cuts, energy-dependent widths or unitarity could be used to improve the parametrization (see, e.g., Ref.~\cite{Doring:2009yv}), but this is not the aim of this study. Note that the background phase $\phi$ and residue phase $\Phi$ are related through coupled-channel unitarity. However, here we fit only one channel in the presence of many other open channels, and
therefore
leave these parameters independent from each other.
Also, if one analyzes lighter channels like $\bar KN$, it is indispensable to include $S$-wave threshold cusps from heavier channels explicitly in the parametrization so that they are not mistakenly identified as resonances. Similarly, thresholds in the complex plane from three-body states may result in false-positive resonance signals~\cite{Ceci:2011ae}. In the analysis of real $K^-p\to K\Xi$ data, a more sophisticated parametrization is employed, see Sec.~\ref{subsec:KXi-Model}.

To avoid that the fit can perfectly reproduce the true solution, the synthetic data were generated using a slightly different parametrization, i.e. including an additional energy dependence in the background i.e. $a\to a+b\,k_f(W)/\Lambda$ for $b\in \mathbb{R}$. From this parametrization, and with realistic choices of free parameters, synthetic data are generated for each partial wave over the same energy range with equal spacing between energy points. Adopting the standard notation $L_{I(2J)}$, four resonances are included in the partial waves $S_{01}$, $P_{11}$, $D_{05}$, and $D_{15}$. The partial waves used to generate the data can be seen in Fig.~\ref{fig:PWA-synthetic}, whereas the data themselves can be seen in Figs.~\ref{fig:Fig2},~\ref{fig:Fig3} and~\ref{fig:Fig32}.

\subsection{Criteria based on information theory}
With the parametrization of Eq.~(\ref{tau}) and synthetic data at hand, we turn to the LASSO method to select the \textit{simplest} model, which describes the data with the minimal number of resonances. In general, the $\chi^2$ is a good measure for determining under-fitting but not over-fitting~\cite{Landay:2016cjw}. Other means to penalize model complexity are needed like the penalization of undesired parameters.
The penalized $\chi_T^2$ is defined as follows
\be
\chi^2_T(\lambda)=\chi^2+P(\lambda)\,,
\label{totalchi}
\ee
where $\chi^2$ denotes the usual measure of the goodness of fit, while the penalty is denoted by $P(\lambda)$ and reads 
\be
P(\lambda)=\lambda^4\sum_{i=1}^{i_{\rm max}} |x_i|\,,
\label{penalty}
\ee
i.e., the $i$-th resonance is penalized through its coupling $x_i$. We allow here for one resonance in each partial wave, i.e., ten resonances altogether, $i_{\rm max}=10$.
In practice, we change (in- or decreasing) $\lambda$ in small steps, minimizing each time $\chi_T^2$. Subsequently, we use various criteria based on information theory in order to determine the optimal $\lambda$. Note also that the power of four in Eq.~(\ref{penalty}) is simply chosen to provide a more convenient graphical representation of these criteria in the following plots. We chose the Bayesian Information Criterion~(BIC) to search for the optimal $\lambda$, defined as
\begin{align}
\text{BIC} &= k_{\rm eff}\log(n) -2\log(L)
\nonumber\\
&= k_{\rm eff}\log(n) + \chi^2+c\,,
\label{eq:BIC}
\end{align}
where $c$ is an irrelevant offset that depends on the number of data but not the model. Here $L$ is the likelihood, $k_{\rm eff}$ denotes the effective number of parameters which changes dynamically as a function of $\lambda$ (see discussion below), while $n$ is the number of data points.
For a normal distributed data, the likelihood can be expressed in terms of the $\chi^2$.

For BIC, the optimal value of $\lambda$ can be determined from the respective minimum. This is because it takes on small values for models with low test error. Note that another common criterion from information theory is the Akaike Information Criterion (AIC). BIC tends to penalize models with more parameters due to the factor $\log (n)$ which allows for a more distinct minimum to be seen and, thus, a clearer indication of which model to favor. The different criteria are compared and illustrated in Ref.~\cite{Landay:2016cjw}. For a further comparison between AIC and BIC, see Refs.~\cite{Tibshirani2011, james2013an}.

The degrees of freedom (d.o.f.) in the penalized fits are increased due to LASSO regularization, which effectively reduces the number of fit parameters. In particular, the d.o.f. are not simply given by the number of data $n$ minus number of parameters $k$ but
\begin{equation}
{\rm d.o.f.}=n-k_\text{eff}\,,
\end{equation}
where $k_\text{eff}$ is the effective number of parameters~\cite{hastie_hastie_tibshirani_friedman_2001},
\begin{equation}
k_{\text{eff}}=\sum_{i=1}^{n} \text{COV}(\hat{y}_i,y_i)\,,
\label{keff}
\end{equation}
given by the covariance of the $i^{\rm th}$ predicted observable $\hat{y_i}$ and the true $i^{\rm th}$ observable  $y_i$.
In practice, we calculate the covariance via bootstrap aggregation, generating $m$ different fits
\begin{equation}
\text{COV}(\hat{y}_i,y_i) = \sum_{j=1}^{m} \frac{(\hat{y}_{i,j} - \bar{\hat{y}}_i)(y_{i,j} - \bar{y}_i)}{m-1}\,,
\end{equation}
where $\hat{y}_{i,j}$ is the $j^{\rm th}$ predicted value for $i^{\rm th}$ data point, and the averaged value for all $m$ predictions for the $i^{\rm th}$ point is denoted by $\bar{\hat{y}}_i= \sum_{j=1}^{m} \hat{y}_{i,j}/m$. The corresponding notation holds for the data points, i.e. $y_{i,j}$ and $\bar y_{j}$.

\begin{figure}
\begin{center}
\includegraphics[width=0.9\linewidth,trim=0.5cm 1cm 0.5cm 0cm]{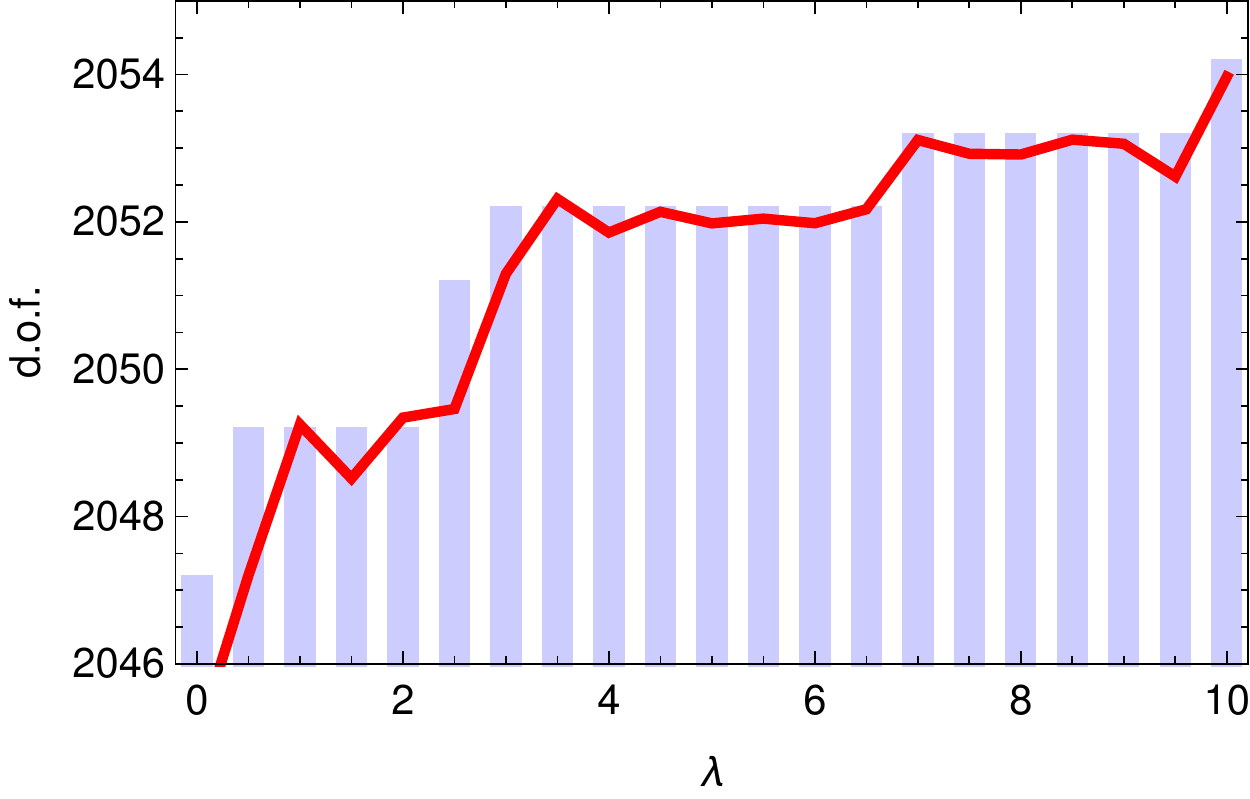}
\end{center}
\caption{%
Degrees of freedom as a function of $\lambda$. The blue bars correspond to counting a parameter $x$ if $|x|>0.01$ while the red line shows the d.o.f. with $k_{\rm eff}$ calculated according to  Eq.~(\ref{keff}).}
\label{fig:Fig4}
\end{figure}

In practical calculations we simplify the described procedure to  determine $k_{\rm eff}$ by counting a fit parameter $x_i$ towards $k_{\rm eff}$ if
it is above some limit, $|x_i|>x_{\rm lim}$. To determine this limit, we perform a simulation with synthetic data and find that $x_{\rm lim}\approx 0.01$, which will be used in the following. The quantity $k_{\rm eff}$ is well determined as can be seen in Fig.~\ref{fig:Fig4}.

\begin{figure*}[ht]
\hspace{1.4cm}\textbf{Forward LASSO}
\hspace{3.1cm}\textbf{Backward LASSO}
\hspace{2.7cm}\textbf{2nd derivative LASSO}\\[-34pt]
~
\begin{center}
\includegraphics[height=.45\linewidth,trim=0.2cm 0.5cm 2cm 1cm]{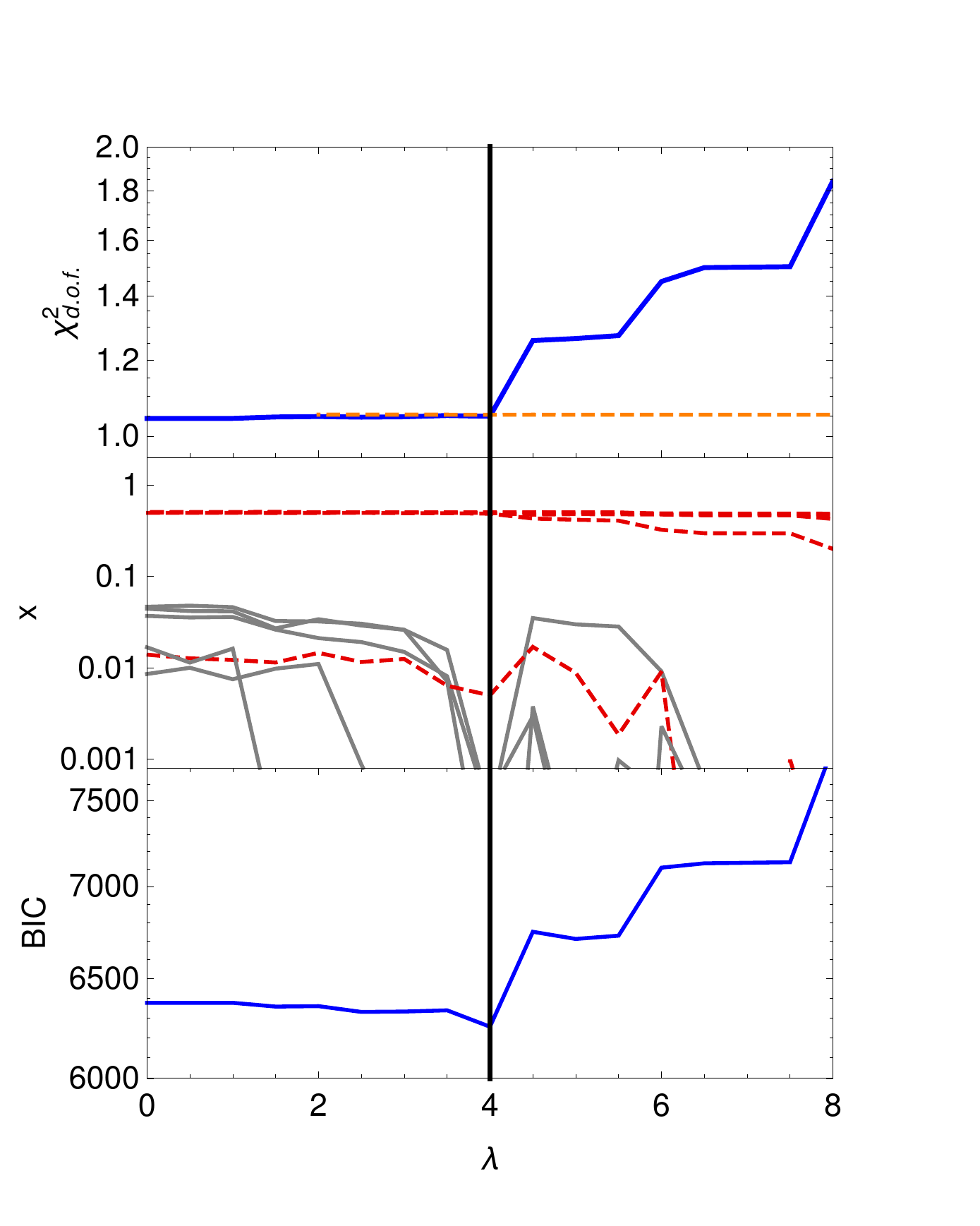}
\includegraphics[height=.45\linewidth,trim=0.4cm 0.5cm 2cm 1cm]{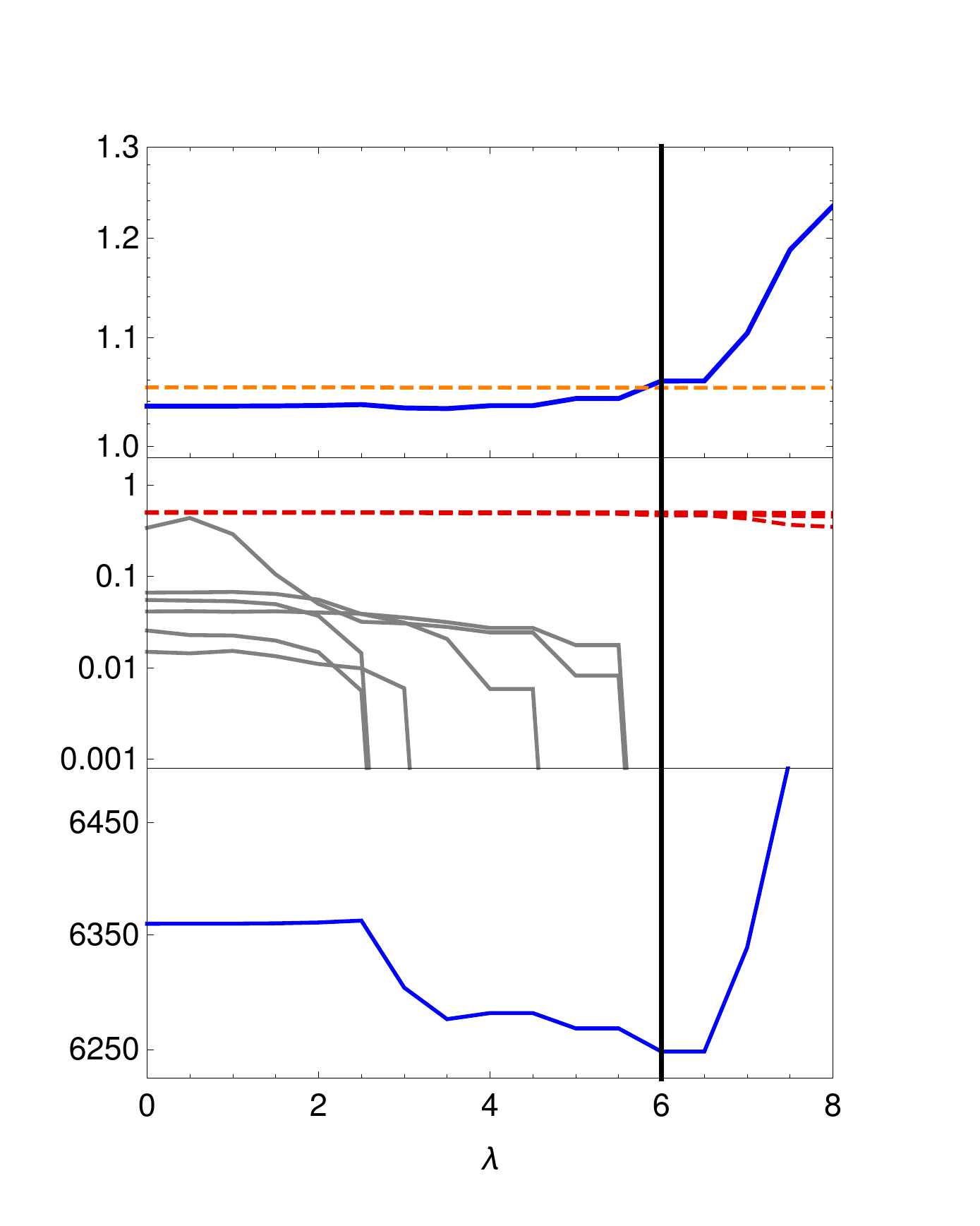}
\includegraphics[height=.45\linewidth,trim=0.4cm 0.5cm 2cm 1cm, clip]{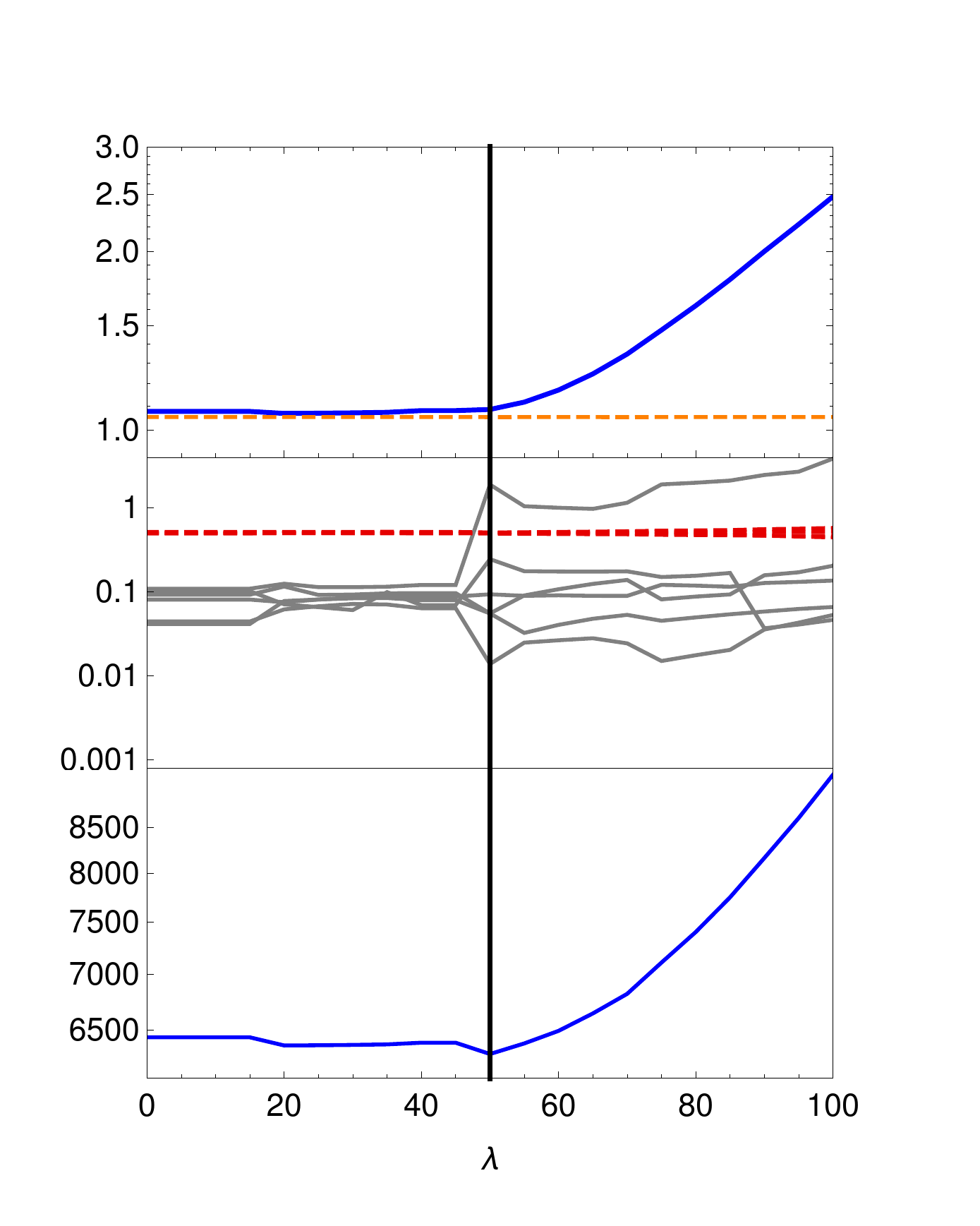}
\caption{Fit results from forward (left), backward automatic shutoff (central) and second-derivative penalty (right) LASSO for the benchmark model. Top panel: $\chi^2$ per degree of freedom in blue with the upper limit of Pearson's $\chi^2$ test per degree of freedom in orange dashed. Middle panel: Absolute value of the ten residues $x$ as a function of $\lambda$ in a logarithmic scale. The red dashed lines indicate the final set of parameters, the gray lines show the unnecessary parameters.
Bottom panel: The Bayesian Information Criterion (BIC). The vertical line signifies the minimum of the BIC that defines the chosen model.}
\label{fig:Fig5}
\end{center}
\end{figure*}

\subsection{LASSO in a benchmark model }
\label{sec:Benchmark Model Fit Results}

In this section, we describe several initial trials using various LASSO implementations on three different benchmark datasets in order to gain a robust understanding of the individual method's strengths and weaknesses before moving onto fitting the real data discussed in detail in Sec.~\ref{sec:real-data}.

The models we use to generate all of the data sets are slightly more complicated than the model we use to fit the data as noted in Sec.~\ref{subsec:parametrization}. All data sets are generated using the same background parameters, energies, and error distributions, but they differ with respect to their resonance content. Our main data set consists of four resonances, each corresponding to a different partial wave with differing masses. However, we also look at a data set containing four resonances, all with the same mass, as well as a set with two groups of two resonances in two different partial waves, all with different masses. In our exploratory analysis, we find, as detailed later in the paper, that some methods are more effective than others, however, we find a consistency among data sets in which particular methods are better than others. In the following we concentrate the discussion on the data set containing four resonances with different masses in four different partial waves. See Sec.~\ref{subsec:parametrization} for details. Our conclusions remain consistent across the other data sets. 

In all three fit strategies discussed in the following, we allow one resonance in each of the ten partial waves ($i_{\text{max}}=10$). 

With our model parametrization, where one parameter being sent to zero ($x_i$) implicitly removes a group of other parameters ($\Gamma_i,\,\Phi_i,\, M_i$), one actually needs to consider the group LASSO ~\cite{hastie_hastie_tibshirani_friedman_2001} instead of the traditional LASSO. The group LASSO can be expressed by the following modification to the penalty term,
\be
P_{gr}(\lambda)=\lambda^4\sum_{i=1}^{i_{\rm max}} \sqrt{p_i} |x_i|\,,
\label{penalty-group}
\ee
where $p_i$ are the number of parameters in the $i^{\rm th}$ group and a group represents a predefined set of variables that are either all included or excluded together. In our case, a group represents the set of parameters which corresponds to the $i^{\rm th}$ resonance. The new term, $p_i$, acts as a weight for various groups, countering the effects caused by potential differences in group size. Here, $p_i = 4, \forall i$ which in practice allows one to absorb them into $\lambda$. In doing so, one retains the same best-fit results as normal LASSO, however, the optimal value of $\lambda$ changes, shifted from the position of the minima of the BIC result. This is an important caveat that must be remembered when using differing resonance parametrizations in various partial waves.

\subsection{Forward LASSO}
\label{sec:fwd}

 For this forward selection model, all ten resonances are initialized with random values selected from Gaussian distributions, i.e. $x_i \sim \mathcal{N}(0,0.25)\,$, $\Gamma_i\sim\mathcal{N}(100,25)\,$, $M_i\sim \mathcal{N}(2500,150)\,$, $\Phi_i\sim\mathcal{N}(0,1)\,$, taking subsequently the absolute value of $x_i$, $\Gamma_i$, $M_i$ to ensure the correct physical scenario. The initialization of the background terms comes from using the fit results from fitting the benchmark model data with no resonances included. We iterate  $\lambda$ stepwise as $10,9.5,...0$, each time minimizing $\chi^2_T$ from Eqs.~(\ref{totalchi}) and (\ref{penalty}), thus penalizing the occurrence of resonances. For each new step in $\lambda$, the converged solution of the previous $\lambda$ is taken as starting value in the fit. In other words, resonances are added until they are all present in the fit, at $\lambda=0$. With BIC we observe a minimum and thus our best model occurs at $\lambda_{\rm opt}=4$; see left panel of Fig.~\ref{fig:Fig5}. This model contains five resonances, the four correct ones and a false one as seen in the same figure (some of the red lines in the figure overlap and are difficult to distinguish). Note also that all of the models from $\lambda=0$ to $\lambda=4$ have a $\chi^2$ within the confidence interval given by a  90\% two-sided confidence level calculated from the $\chi^2$ distribution (referred to as ``Pearson's $\chi^2$ test'' in the following). While the best fit results for the forward model is not in complete agreement with the benchmark model, it still represents a good local minimum in $\chi^2$ and a starting point for initial guesses of subsequent optimizations.

\subsection{Backward Automatic Shutoff LASSO }

In linear regression one can expect that the LASSO path, i.e., the estimated parameters as function of $\lambda$ in parameter space, does not depend whether forward selection or backward selection is applied. There is only one local minimum and the $\chi^2$ is a multi-dimensional parabola in parameter space. Our current problem, however, is inherently non-linear because the observables are bilinear in the parameters (c.f. App.~\ref{sec:appA})

In particular, there are multiple local minima and the result of the backward selection (starting with $\lambda=0$ and dynamically updating the initial values as described in the previous section) depends on the local minimum one starts from. 

In the backward selection, we start with the minimum determined at $\lambda=0$ with the forward selection discussed before. As before, we iterate $\lambda$ in steps at values $0,0.5,\ldots,10$, each time minimizing $\chi^2_T$ from Eqs.~(\ref{totalchi}-\ref{penalty}) for $i_{\text{max}}=10$ and updating the initialization of each fit by the converged fit of the previous value of $\lambda$. As a result the minimum in BIC occurs at $\lambda_{\rm opt}=6$ at which the resonances are correctly selected and their properties are very close to their correct values (masses, widths, couplings).

Next, we discuss a {\it greedier} version of the backward selection, referred to as {\it backward automatic shutoff} in the following. The modification is that once a pole residuum $x_i$ becomes smaller than $10^{-3}$, which is our shutoff criterion, that resonance is permanently removed from the model and is no longer fit for the remaining iterations. From BIC results shown in the central panel of Fig.~\ref{fig:Fig5} we see the minimum, and thus our best model, occurs at $\lambda=6$. This model contains only the four genuine resonances, successfully sending all of the other resonance couplings to zero as shown. The minimum in BIC also coincides  with the intersection of the $\chi^2$ with the value given from Pearson's $\chi^2$ test indicating that the model passes the test.

\subsection{Second-Derivative Penalty}
\label{SecDeriv}

In many approaches to extract the baryon spectrum, it is not possible to directly penalize the size of the resonance residues as tested before. In dynamical coupled-channel approaches one can still penalize bare resonance couplings and, thus, remove the dressed resonance poles. Yet, in these approaches, the non-linear meson-baryon dynamics can lead to the formation of resonance poles~\cite{Doring:2009bi}, and it is difficult to pin down \textit{the} corresponding parameters responsible for resonance formation. In the SAID approach~\cite{Workman:2012jf} resonances are almost exclusively generated through the unitary coupled-channels dynamics if required by data. 

One way of minimizing the number of resonances, when fit parameters cannot be clearly attributed to their existence, is to penalize the second derivative of the partial-wave amplitude. 

In this study, we are working in a one-channel approximation, with no prominent two-body threshold opening above $K\Xi$  such that non-analyticities for physical energies are assumed to be negligible. Accordingly,  we introduce the penalty
\begin{align}
\label{eq:penaly2}
P(\lambda)= \lambda^5\sum_{i=1}^{10} &\frac{\int_{m_K+m_{\Xi}}^{W_{\rm max}}\left|\frac{\partial^2}{\partial W^2} \tau_i(W)\right|^2\, dW}{\int_{m_K+m_{\Xi}}^{W_{\rm max}} |\tau_i(W)|^2\, dW }\,,
\end{align}
where index $i$ denotes the corresponding partial wave indices $(I,J,L)$, and $W_{\rm max}=3200$~MeV is the maximum energy of the data. For numerical convenience, we penalize here only the resonance term in $\tau$, i.e., the second term in Eq.~(\ref{tau}).

\begin{figure}
\begin{center}
\includegraphics[width=.99\linewidth]{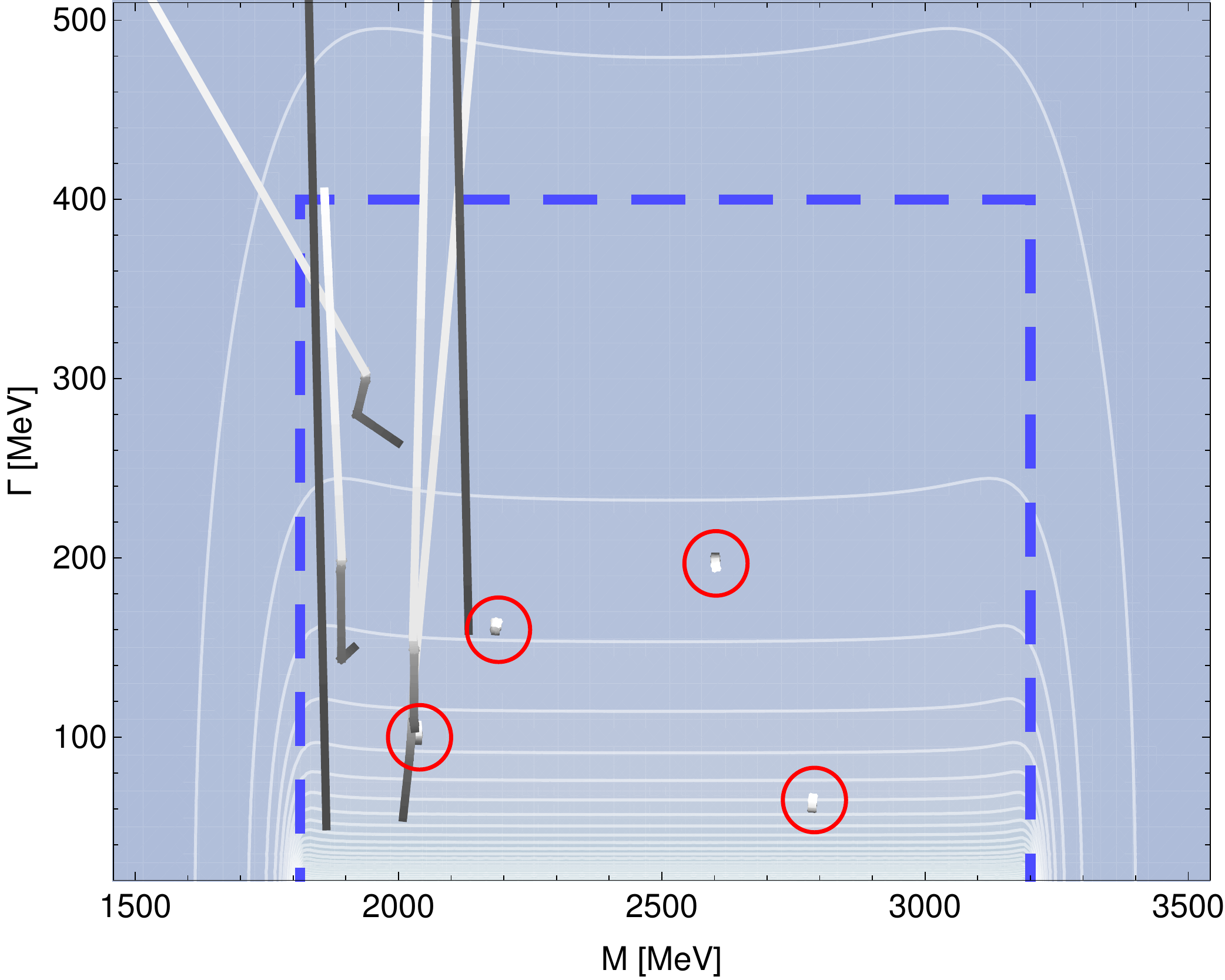}
\caption{Derivative penalty. Resonance trajectories (thick solid lines) in the $(M,\Gamma)$ plane as function of penalty parameter $\lambda$ from $\lambda=0$ (dark shading) to $\lambda=\lambda_{\rm opt}$ (light shading). The very short trajectories of significant resonances are highlighted by red circles. The thick blue dashed line shows the $(M,\Gamma)$ region in which resonance parameters are counted towards the total number of parameters. The typical penalty size is indicated with white contours ranging from large penalty (close to $\Gamma=0$) to small penalty (large $\Gamma$ and/or high/low $M$.)
}
\label{fig:trajectories}
\end{center}
\end{figure}

The introduced penalty term is significantly different from the previous one of penalizing the resonance couplings $|x|$. This allows for resonances to effectively disappear by their widths becoming so large that they flatten out and become indistinguishable from the background, or by their masses moving outside of the fitted region. The typical form for this penalty is indicated in Fig.~\ref{fig:trajectories} with the white contours ranging from large penalty (close to the physical axis) to small penalty (for wide and/or sub/above-threshold resonances.)

For the determination of the resonance spectrum, we proceed like in case of backward LASSO, i.e., from the same local minimum at $\lambda=0$, dynamically updating $\lambda$. With respect to counting parameters to determine the degrees of freedom, the four parameters of a given resonance are only counted in BIC if the resonance pole is within a certain $(M,\Gamma)$ region. This ``resonance area'' is indicated in Fig.~\ref{fig:trajectories} with the thick blue dashed line. The window in mass reaches from threshold to $W_{\rm max}$, given by the maximum energy of available data, and in width up to $\Gamma_{\rm max}$. The $\chi^2$ and BIC are shown in Fig.~\ref{fig:Fig5} in the right-hand panel. The minimum in BIC occurs at $\lambda=\lambda_{\rm opt}\approx 50$ which coincides with one false resonance leaving the resonance area (see Fig.~\ref{fig:trajectories} at around $(M,\Gamma)=(1.85,0.4)$~GeV). At $\lambda=\lambda_{\rm opt}$, the significant resonances have barely moved (short trajectories highlighted by red circles) while the false resonances are completely driven out of the resonance area.

We have checked explicitly that for  {${\Gamma_{\rm max}\in [250,400]}$~MeV} different values of $\lambda_{\rm opt}$ are obtained, but in each case leading to the same best resonance content. As for backward LASSO, the second-derivative penalty is able to correctly identity the four genuine resonances while eliminating the others by sending their widths above $\Gamma_{\rm max}$ and/or their masses out of the fitted energy window.

\subsection{Discussion}

The discussed LASSO variants perform similarly. Backward LASSO and second-derivative penalty are able to correctly identify which resonances are present in the data while the forward selection is off by one resonance. The automatic shutoff method leads to a more pronounced minimum in the BIC than the second-derivative penalty. It is, however, greedier in the sense that once a parameter is zero, it is forever removed from the fit. This can become an issue if there are multiple local minima and the fit cannot explore them because parameters have been shut off.

The second-derivative penalty has the advantage that parameters are not removed at all, but they can still contribute to shape the background that varies slowly with energy. This possibly protects the fit against bias in the background terms: In case the background parametrization is not flexible enough this could lead to false-positive resonance signals. 

Yet, the derivative penalty has a slightly different meaning than the penalty of Eq.~(\ref{penalty}). While in the latter, resonance poles are completely removed from the partial-wave amplitude, the derivative penalty moves resonance poles far away from the physical axis and the region of fitted data. From a phenomenological point of view, these scenarios are quite similar to each other. However, if spectra from theory are to be tested with phenomenology, wide resonances pose a problem because in quark models and related approaches, resonance widths cannot be reliably determined and one does not know if a pole in the complex plane far away from the real axis corresponds to a quark-model state or not. Such questions are, however, not of interest for this data-driven phenomenological approach.

Higher derivatives in the penalization are also possible and, if they can be reliably evaluated, even desirable: For example, if one has a small resonance signal on top of a large background,  the denominator of Eq. (\ref{eq:penaly2}) could become large and the penalty small. Replacing both the numerator and denominator with higher derivatives might be more suitable to detect such special circumstances.

The obvious disadvantages of the derivative penalty lie in the more complicated analytic structure in form of threshold cusps  in the physical scattering region on or close to the real axis~\cite{Ceci:2011ae}. In the analysis of the $K^-p\to K\Xi$ reaction, we assume that those thresholds (e.g., from $K^*\Xi$ or $K\Xi^*$) play no role. One could explicitly exclude threshold regions from the integrals of Eq.~(\ref{eq:penaly2}) but then has to pay attention to resonances on hidden sheets that might enhance thresholds.

Another possibility to penalize resonances close to the physical axis, not explored here, is given by suitable closed-contour integrals on the unphysical Riemann sheets~\cite{Doring:2009yv} that could be used to penalize the size of resonance residues. This method can deal with threshold openings if the contour is chosen appropriately but would fail if residues of two or more resonances cancel. 

Due to its performance identifying correct resonance content (of synthetic data) and its simplicity, the backward automatic shutoff LASSO will be used in the next section for the determination of the resonance spectrum with actual data from experiment.

\section{\boldmath Analysis of $\bar{K} N \to K\Xi$ with LASSO}
\label{sec:real-data}

In this section, we present a blindfold analysis of the resonance content of the actual data using LASSO in combination with BIC as explained in the previous sections, see also Refs.~\cite{TS96,hastie_hastie_tibshirani_friedman_2001,james2013an} and \cite{S78}. To this end we consider the reaction $K^-p\to K \Xi$. This choice of the reaction is on purpose for this exploratory calculation because for many existing data one often encounters a situation where data sets from different experiments are inconsistent with each other due to underestimation of systematic uncertainties. Also, some experimental data sets are of very poor quality, which makes the extraction of resonances from such data difficult. The $K^-p\to K\Xi$ reaction is chosen here to test LASSO for its robustness against such a database. 

Obviously, the resonance content extracted from the data can depend on which data sets one includes in the analysis. Thus, in general, the selection of the data to be considered is the first step toward an extraction of resonances. To this end, here, we apply the so-called self-consistent $3\sigma$ criterion \cite{Perez:2013jpa,Perez:2014yla}. Once the data sets to be included in the analysis are selected, we proceed to fit the model parameters using the LASSO method in combination with BIC (LASSO+BIC). Our model for the reaction at hand contains initially all the known above-threshold hyperon resonances from the Particle Data Group (PDG)~\cite{PDG16}, irrespective of their rating status. The LASSO+BIC method will tell us which resonances will actually be required to fit the data. 

\subsection{Calculation of the merit function}
\label{subsec:KXi-MerFunc}

\begin{table*}[thb]
\centering
\begin{tabular}{c@{\extracolsep{1.65em}}S[table-format=1.1]S[table-format=3.0]cS[table-format= 1.1]|cS[table-format=4.0]S[table-format =1.1]cS[table-format= 1.1]} \hline
\multicolumn{4}{c}{$\Lambda$ states} & & \multicolumn{4}{c}{$\Sigma$ states} \\
\hline State & {$m_R$ (MeV)} & {$\Gamma_R$ (MeV)} & Rating & &
 State & {$m_R$ (MeV)} & {$\Gamma_R$ (MeV)} & Rating  \\ \hline
$\Lambda(1810)$ $1/2^+$ & 1810  & 150 & *** & &
  $\Sigma(1840)$ $3/2^+$ & 1840  & 100 & * \\
$\Lambda(1820)$ $5/2^+$ & 1820  & 80 & ****  & &  
  $\Sigma(1880)$ $1/2^+$ & 1880  & 194 & **  \\
$\Lambda(1830)$ $5/2^-$ & 1830  & 95 & **** & &
  $\Sigma(1900)$ $1/2^-$ & 1900  & 191 & *  \\
$\Lambda(1890)$ $3/2^+$ & 1890  & 100 & **** & & 
$\Sigma(1915)$ $5/2^+$ & 1915  & 120 & ****  \\
$\Lambda(2000)$ $\ \ \ \ ?^?$ & 2000  & 167 & * & & 
  $\Sigma(1940)$ $3/2^+$ & 1941  & 400 & *  \\
$\Lambda(2020)$ $7/2^+$ & 2020  & 195 & * & & 
  $\Sigma(1940)$ $3/2^-$ & 1940  & 220 & ***  \\
$\Lambda(2100)$ $7/2^-$ & 2100  & 200 & **** &  & 
  $\Sigma(2000)$ $1/2^-$ & 2000  & 273 & *  \\
$\Lambda(2110)$ $5/2^+$ & 2110  & 200 & *** & &
  $\Sigma(2030)$ $7/2^+$ & 2030  & 180 & ****  \\
$\Lambda(2325)$ $3/2^-$ & 2325  & 169 & * &  &
  $\Sigma(2070)$ $5/2^+$ & 2070 & 220 & *  \\
& & & & &  
  $\Sigma(2080)$ $3/2^+$ & 2080 & 177 & **  \\
& & & & &  
  $\Sigma(2100)$ $7/2^-$ & 2100 & 103 & *  \\
& & & & &  
$\Sigma(2250)$ $\ \ \ \ ?^?$ & 2265 & 100 & ***   \\
\hline
\end{tabular}
\caption{$\Lambda$ and $\Sigma$ hyperons considered in this work. Masses ($m_R$) and widths ($\Gamma_R$) are extracted from the PDG~\cite{PDG16}, except for the $\Sigma(2250)$ resonance, whose mass has been adjusted to reproduce the peak position of the bump structure seen in the total cross section data. For one- and two-star resonances, where no estimates are available, we take the average of the values quoted in PDG. In this average for the width of the $\Sigma(2070)5/2^+$, we have excluded the 906 MeV width by Kane \cite{Kane:1972qa}.} 
\label{tbl:hyperons}
\end{table*}
 
In general, the theoretical description of a given experimental data set is achieved by fitting the model parameters through a minimization procedure of the 
merit function 
\begin{equation}
\chi^2 = \sum_{k} \chi^2_k\,, 
\label{eq:E1}
\end{equation}
where the summation runs over all datasets, specified by the index $k$. For each dataset $k$, $\chi^2_k$ is given by  
\begin{equation}
\chi^2_k = \sum_{i=1}^n \left(\frac{y_i - Z_k\hat y_i}{\delta y_i} \right)^2 
+ \left(\frac{Z_k - 1}{\delta_{{\rm sys}\, k}} \right)^2\,,
\label{eq:E1a}
\end{equation}
where, $y_i$ and $\delta y_i$ are, respectively, the experimental value and corresponding statistical uncertainty of the observable at the kinematical point (total energy and scattering angle) specified by the index $i$. The number of data points in each data set is denoted by $n$, while $\hat y_i$ stands for the model fit value for that observable. The contribution to $\chi^2_k$ arising from systematic uncertainties is addressed by the last term in the above equation, expressed by the systematic uncertainty ($\delta_{{\rm sys}\, k}$) and the scaling factor ($Z_k$). 
 We note that every experimental data set can be subject to a known and common systematic uncertainty (normalized data), an arbitrarily large systematic uncertainty (floated data) or no systematic uncertainty at all (absolute data). 
Absolute data have $\delta_{{\rm sys}\, k} = 0$ and are not scaled ($Z_k = 1$). The correct value of $Z_k$ for normalized and floated data is obtained by minimizing $\chi^2_k$ with respect to $Z_k$. This leads to 
\begin{equation}
Z_k = 
\left(
\sum_{i=1}^n \frac{y_i\, \hat y_i}{\delta y_i^2} + \frac{1}{\delta_{\text{sys}\, k}^2} 
\right)    
\Bigg[ 
\sum_{i=1}^n \Bigg(\frac{\hat y_i}{\delta y_i}\Bigg)^2 + \frac{1}{\delta^2_{\text{sys}\, k}} \Bigg]^{-1}\,.
\label{eq:E2}
\end{equation}

Due to the nature of the currently available data for $K^-p \to K \Xi$, as discussed in the following subsection, where systematic uncertainties are unknown, we treat the data as absolute, i.e., set $\delta_{{\rm sys}\, k}=0$ and $Z_k=1$ in this work. This is also what was done in Ref.~\cite{Jackson:2015dva}. Furthermore, each data point is considered to be a data set of its own, i.e., $n=1$.
The total $\chi^2$ given by Eq.~(\ref{eq:E1}) is then minimized using the MINUIT minimization code.  
As systematic uncertainties are neglected, problems tied to the d'Agostini bias~\cite{DAgostini:1993arp, Ball:2009qv} play no role.

\begin{figure*}[t] \centering
\includegraphics[height=0.35\textwidth,clip=1]{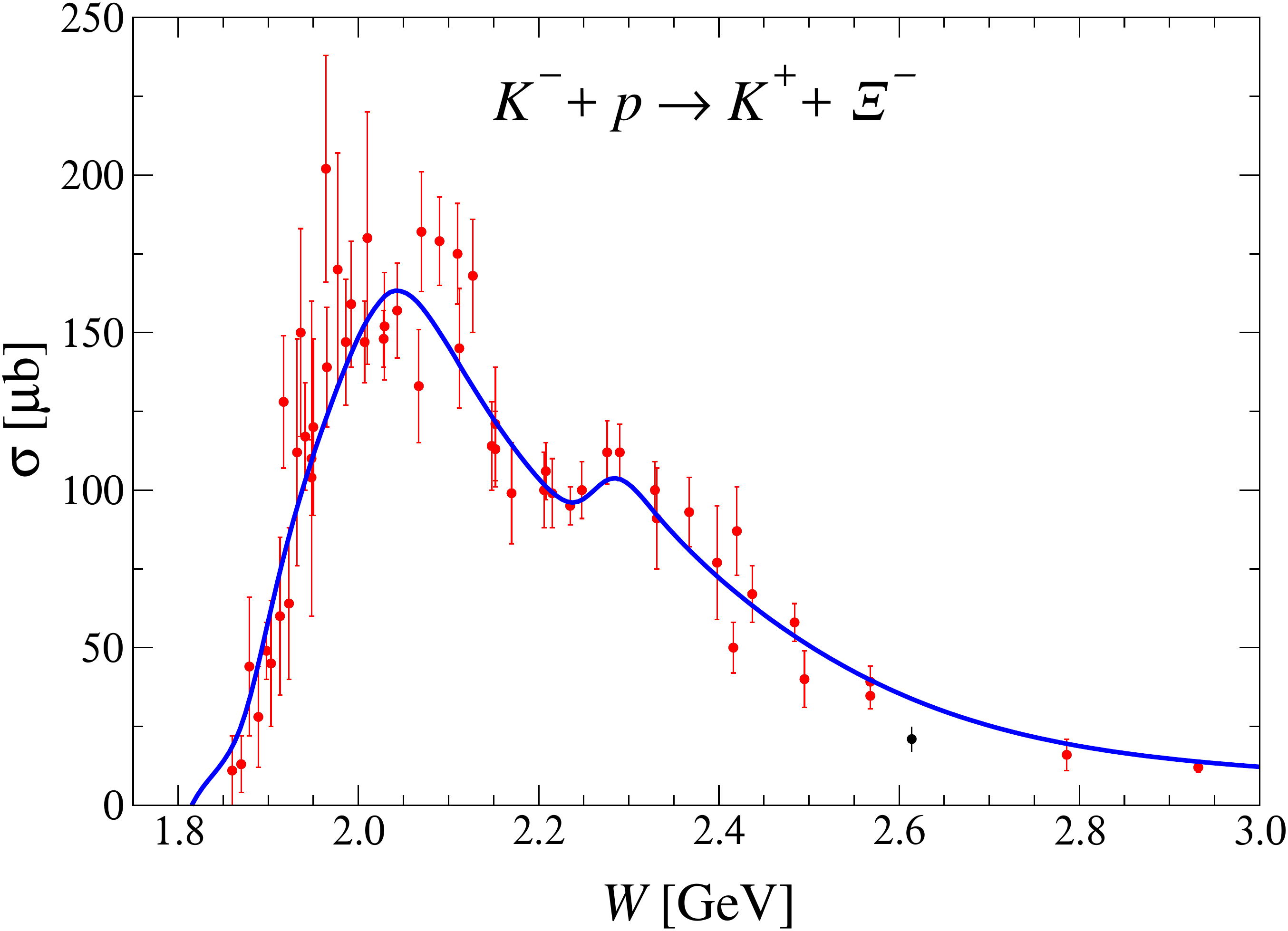}  
\includegraphics[height=0.35\textwidth,clip=1]{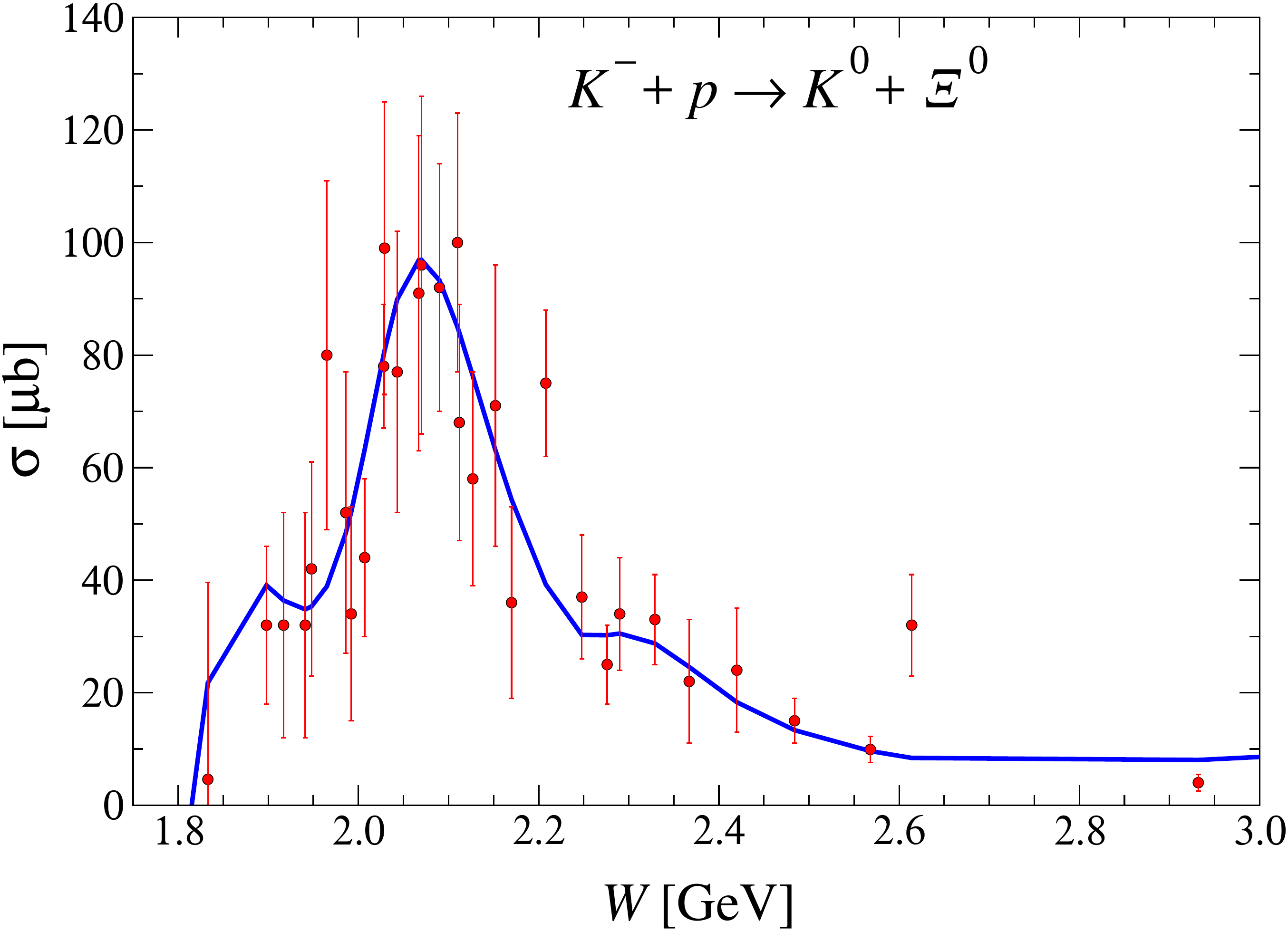}  \qquad
\includegraphics[height=0.32\textwidth,clip=1]{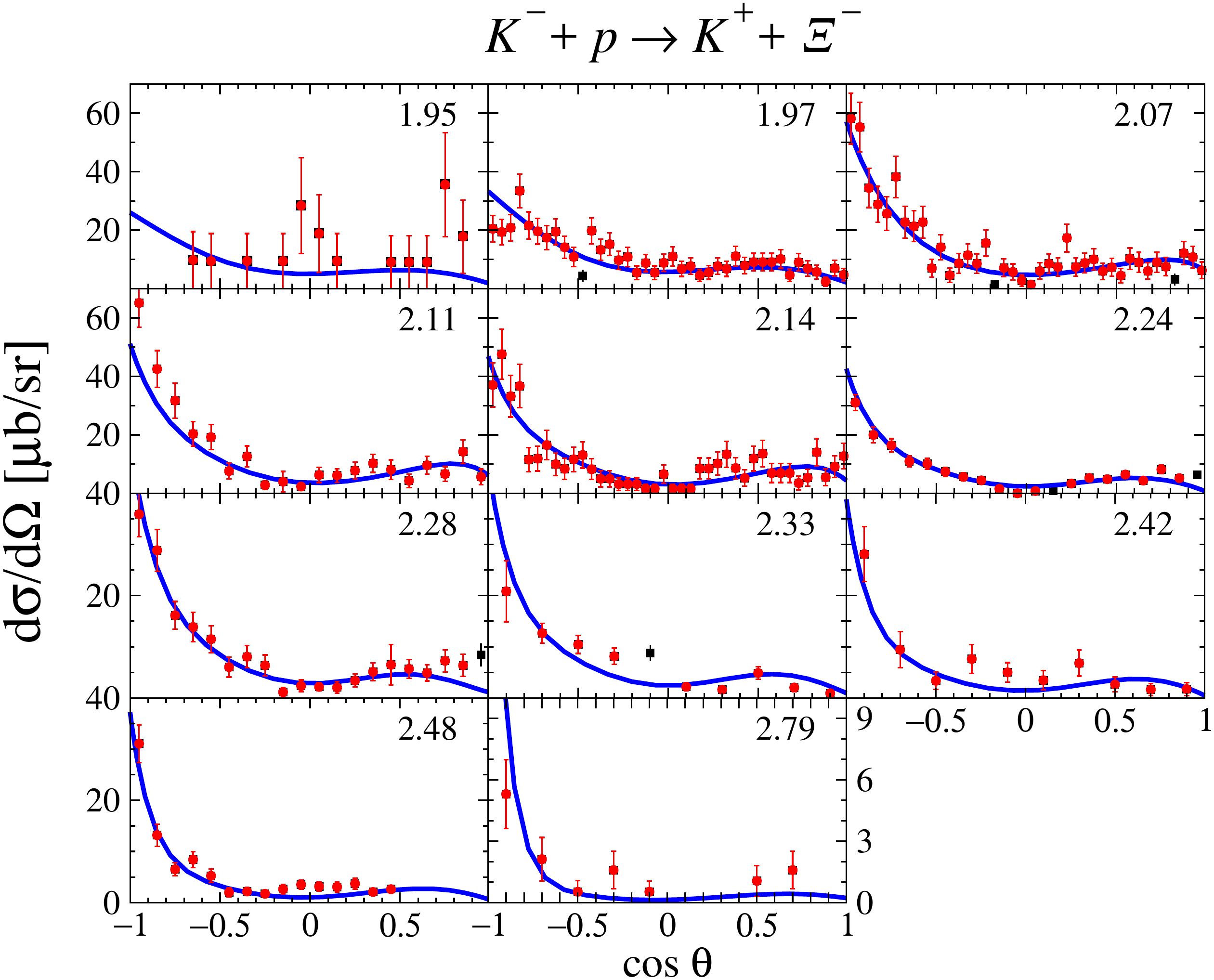}  
\includegraphics[height=0.32\textwidth,clip=1]{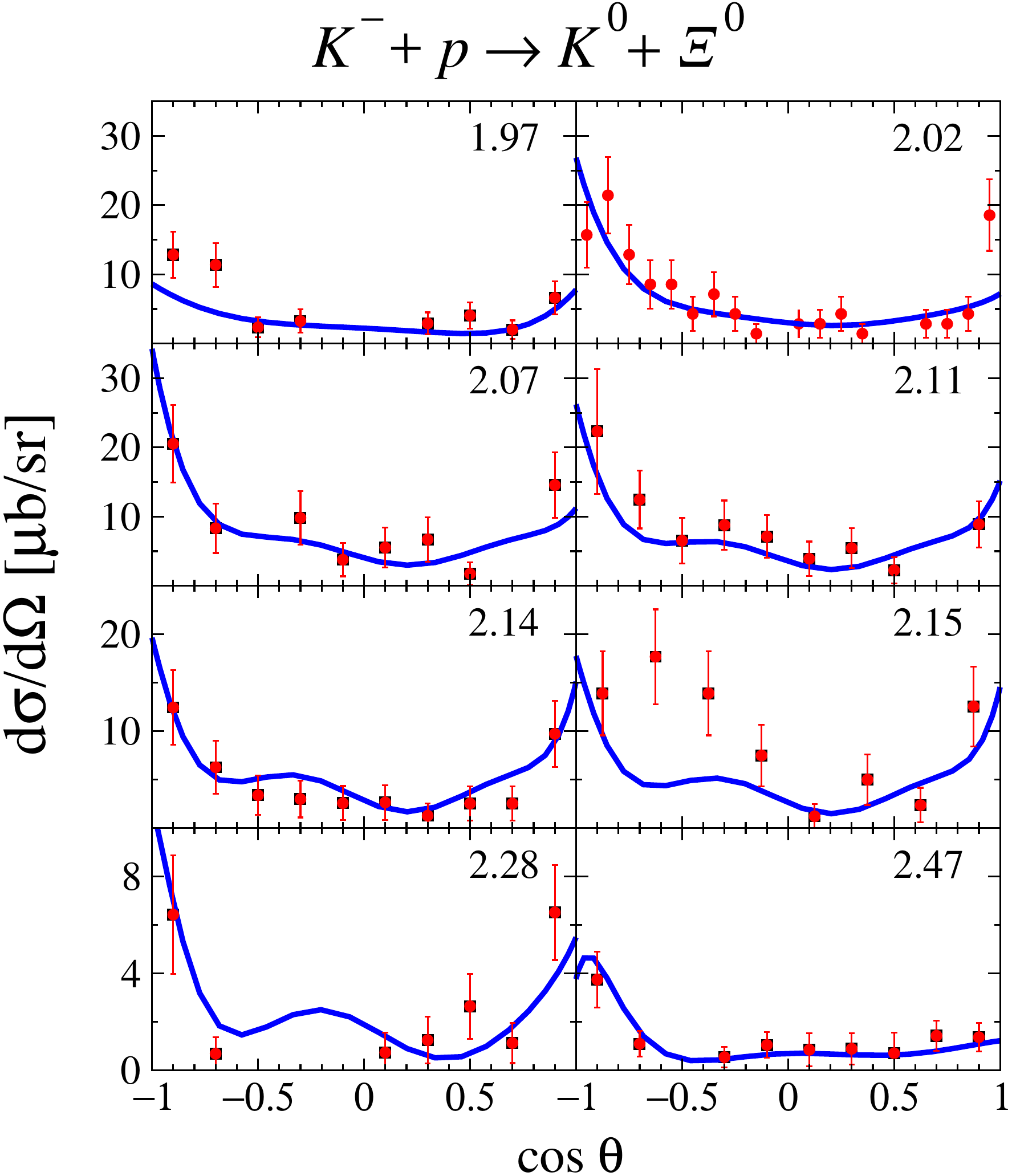}  
\includegraphics[height=0.32\textwidth,clip=1]{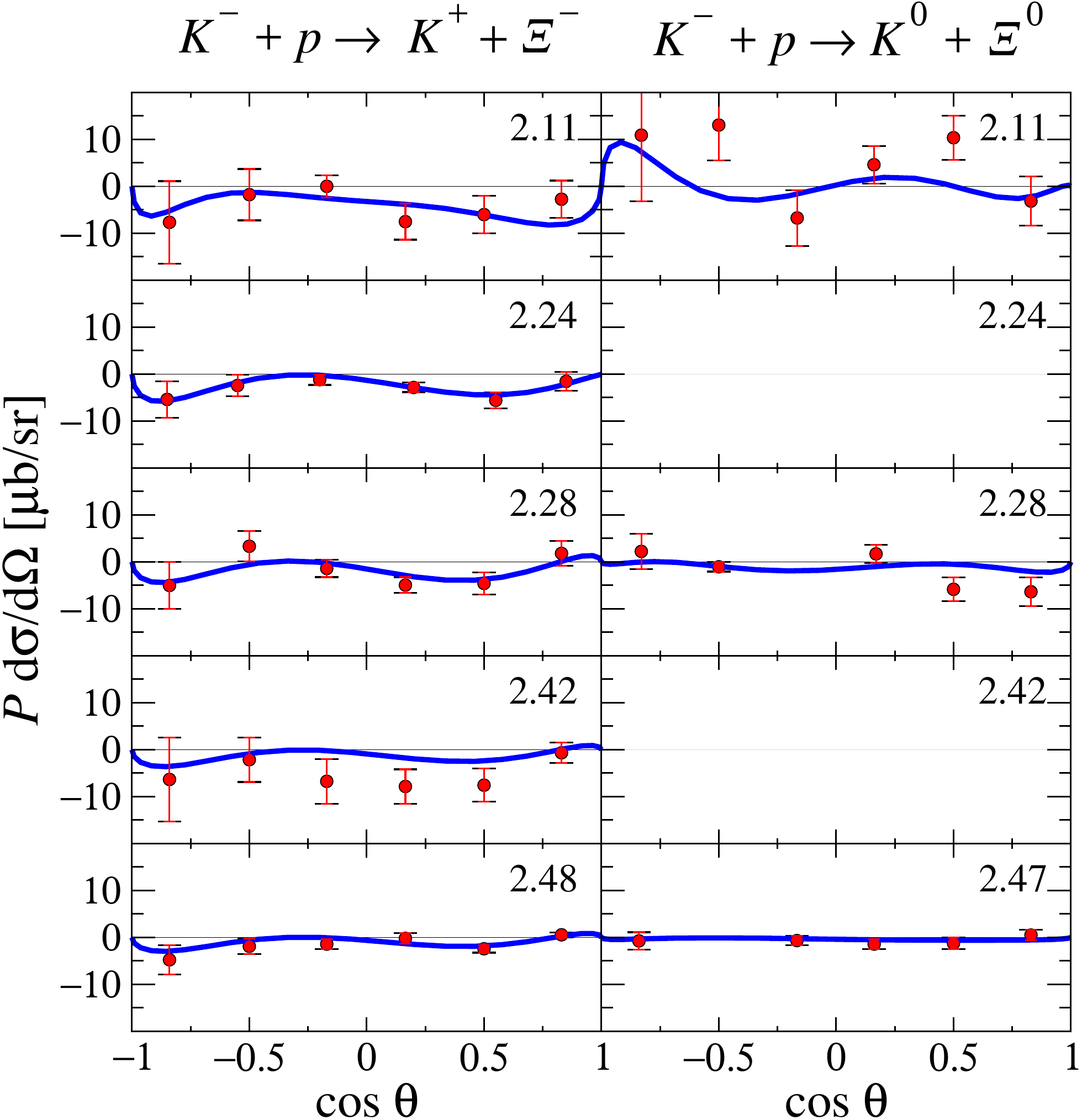}  
\caption{Full unpruned data (black+red) and $3\sigma$-pruned data (red) as described in the text. The fit results using the over-parametrized model described in Sec.~\ref{subsec:KXi-DataSelect} are shown by the blue curve with $\chi^2_{d.o.f.} = 2.53$ with respect to the full data. The numbers in the plots of $d\sigma/d\Omega$ and $P$ indicate the total scattering energy $W$ in GeV.}
\label{fig:KXi-3sigma}
\end{figure*}

\subsection{Data selection} 
\label{subsec:KXi-DataSelect}
The reaction process $K^-p \to K \Xi$ has been studied experimentally, mainly, throughout the 1960's~%
\cite{Pjerrou:1962pc,Carmony:1964zza,Berge:1966zz,Haque:1966mda,London:1966zz,Trippe:1967wat,Trower:1968zz,Merrill:1968zz,Burgun:1969ee,Dauber:1969hg}, which was followed by several measurements made in the 1970's and 1980's~%
\cite{Scheuer:1971zw,DeBellefon:1972gm,Carlson:1973td,Rader:1973ja,Griselin:1975pa,Briefel:1977bp,Dumbrajs:1983jd}. The existing data (total cross sections, differential cross sections, and recoil polarization asymmetries) are rather limited and suffer from large uncertainties. The total cross section and some of the differential cross-section data are tabulated in 
Ref.~\cite{FMMR83b}. Some of them are not in tabular (numerical) form that can be readily used but are given only 
in graphical form or as parametrization in terms of their Legendre polynomial expansions. 
In Ref.~\cite{Sharov:2011xq}, Sharov \textit{et al.} have carefully considered the data extraction from these papers. We have checked that the extracted data are consistent with those in the original papers within the permitted accuracy of the check. In the present work, we use these data from Ref.~\cite{Sharov:2011xq}. No differential cross sections given in terms of the Legendre polynomial expansions are included.

From the database mentioned above we select the data points to be included in our analysis using the self-consistent $3\sigma$ criterion applied in Refs.~\cite{Perez:2013jpa,Perez:2014yla} to the potential-model analyses of $NN$ scattering. This is an improved version of the $3\sigma$ criterion introduced by the Nijmegen group in their 1993 partial-wave analysis \cite{Stoks:1993tb} which became an essential aspect of their success and the subsequent high-quality fits of the $NN$ scattering data \cite{Stoks:1994wp,Wiringa:1994wb,Machleidt:2000ge,Gross:2008ps}. This criterion discards mutually incompatible data, but can also prevent a fraction of the data to contribute to the final fit. This is so because no distinction is made between mutually incompatible data sets in similar kinematical conditions and which of them, if any, are actually incompatible with the remaining data in different kinematical conditions. The latter is encoded in the phenomenological parametrization which links all kinematical regions. The self-consistent $3\sigma$ criterion is an extension of the $3\sigma$ criterion, which differentiates both situations.

For a set of $n$ measurements with Gaussian distribution, the quantity $z \equiv \chi^2_k/n$ follows a re-scaled, re-normalized $\chi^2$ distribution,
\begin{equation}
{\cal P}_n(z) = \frac{n(nz/2)^{n/2 - 1}}{2\Gamma(n/2)}\, e^{-nz/2}\,.
\label{eq:DP1}
\end{equation}
Here, $\Gamma(x)$ stands for the usual gamma function.
According to the $3\sigma$ criterion, a dataset (here: a single data point) is considered inconsistent with the rest of the database if its statistics $z>z_{\rm max}$ where $z_{\rm max}$ is given by the cumulative distribution function, CDF$[{\cal P}_n(z_{\rm max})]=1-0.0027$. In most cases, a dataset will have a highly improbable $z$-value if the systematic errors are underestimated ($z$ will be very large). The discussed one-sided criterion reads
\begin{align}
\int_{z_{max}(n)}^\infty  {\cal P}_n(z)\, dz & = \frac{\Gamma(n/2, nz_{max}/2)}{\Gamma(n/2)} = 0.0027\,,
\label{eq:DP2}
\end{align}
where $\Gamma(x,y)$ is the incomplete gamma function.
One could also consider a two-sided criterion as in Ref.~\cite{Perez:2014yla} to exclude data with too good of a $\chi^2$. However, in the present situation, in which every data point counts as a data set, this does not make much sense; there is no problem if the $\chi^2$ of a single point is very small; the problem arises only if the $\chi^2$ of an entire data set is too small, and then one might conclude that the errors in that data set are overestimated and a two-sided criterion might be needed. In a test, we found no evidence for overestimated error bars that would justify the usage of a two-tailed pruning criterion.

In practice, the above methodology is applied as follows: 1) we fit the entire database (unpruned data)  with some phenomenological model to 
represent the database. The model used just in this subsection for data pruning purposes is chosen to be over-flexible in the sense that the pruning should not occur due to a biased parametrization. This model is constructed based on the model of Ref.~\cite{Jackson:2015dva}. The differences are that, here, we include more contact and resonance contributions. In addition, we relax the constraints imposed in Ref.~\cite{Jackson:2015dva} on the complex phases in the contact amplitudes as well as the constancy of the masses and widths of the resonances. All these differences make the model more flexible. As to the additional number of resonances included, we have made sure that these does not start to fit the obvious statistical fluctuations in the data.
2) Using the fitted model, we calculate $z$ of each data point, subsequently pruning the database according to the $3\sigma$ criterion described above. 3) The pruned database is then fitted anew and the $3\sigma$ criterion is applied again to the entire unpruned database to obtain a new pruned database. The process is repeated until  self-consistency is reached, i.e, the pruned database remains unchanged after the iterations. 

The results of the pruning according to the self-consistent $3\sigma$ criterion described above are shown in Fig.~\ref{fig:KXi-3sigma}. Only 10 data points out of 448 in total are outside the allowed range of $z$.

\subsection{Theoretical model}
\label{subsec:KXi-Model}

In the analysis of the $\bar{K} N \to K \Xi$ reaction we use the theoretical model of Ref.~\cite{Jackson:2015dva}, except for the above-threshold resonances considered. In contrast to Ref.~\cite{Jackson:2015dva}, in the present blindfold analysis, we consider all the above-threshold hyperon resonances, irrespective of their PDG rating status. Furthermore, the PDG \cite{PDG16} does not assign the spin-parity quantum numbers for the $\Sigma(2250)$ and $\Lambda(2000)$ resonances. The analyses of Ref.~\cite{DeBellefon:1972gm}  provide two possible parameter sets for the $\Sigma(2250)$, one with $J^P = 5/2^-$ at about $2270 \pm 50$~MeV and another
one with $J^P = 9/2^-$ at about $2210 \pm 30$~MeV. In the present work, we assume the $\Sigma(2250)$ to have $J^P = 5/2^-$ with a mass of $2265$~MeV, the primary reason being that the total cross section in $K^-p\to K^+\Xi^-$ shows a small bump structure at around $2300$~MeV, which is well reproduced in our model with these parameter values. 
We refer to this resonance as $\Sigma(2265)5/2^-$ in the following.
For the $\Lambda(2000)$ resonance, we adopt $J^P=1/2^-$, the only  quantum numbers claimed in Ref.~\cite{Zhang:2013sva}. The PDG also quotes a one-star $\Lambda(2050)3/2^-$ resonance with a mass of $2056$~MeV and width of $493$~MeV.
We do not consider this resonance in our study here due to its width being larger than the maximum value of 400 MeV adopted in the present work (cf.~Sec.~\ref{SecDeriv}). 
Neither do we consider the high-spin three-star $\Lambda(2350)9/2^-$ resonance. 
The inclusion of baryon resonances requires the knowledge of the corresponding propagators and transition vertices, which is not a trivial task both conceptually and numerically, especially, for high-spin resonances. Indeed, it is well known that, unlike for spin-1/2 resonances, the construction of propagators for higher-spin resonances is not a straightforward procedure.    In principle, the propagators and transition vertices for high-spin resonances can be obtained, e.g., following the Rarita-Schwinger approach \cite{BF57,R66,CHANG67}. In fact this is the case for spin-5/2 and -7/2 resonances which have been tested and applied in the description of many reactions \cite{NOH11,MON11,Jackson:2015dva,Wang:2017tpe} and also used in the present work. However, to our knowledge, spin-9/2 resonances have never been considered in microscopic calculations where the full Dirac-Lorentz structures of the corresponding propagators and vertices are required. Furthermore, the number of Lorentz indices to be contracted in evaluating the reaction amplitude involving baryon resonances in the intermediate states increase with the resonances' spin. In fact, for a resonance with spin-$j$, its propagator has $(2j-1)$ Lorentz indices and its transition vertex, $(j-1/2)$ indices. Hence, the number of Lorentz indices to be contracted increases rapidly with the spin of the resonance, making the evaluation of the reaction amplitude containing the high-spin resonances, such as $\Lambda(2350)9/2^-$, very much time consuming. Thus, for the reasons given above, the inclusion of spin-9/2 resonances is beyond the scope of the present work. The full set of  resonances considered in the present work is listed in Tab.~\ref{tbl:hyperons}.

\begin{figure}[t]
\includegraphics[width=0.9\linewidth,clip=1]{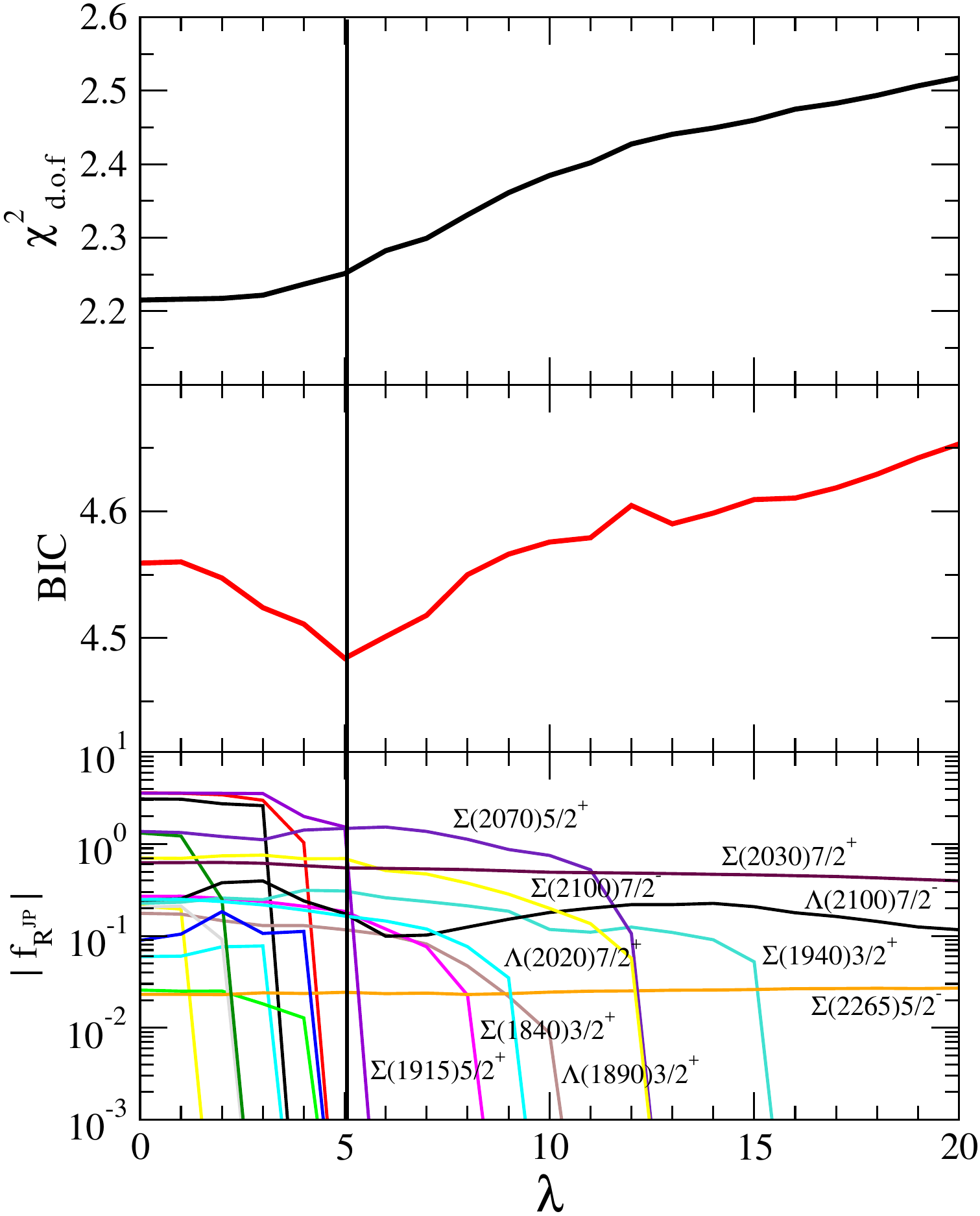}
\caption{$\chi^2_{\rm d.o.f.}$ (upper panel), BIC (middle panel), and the absolute value of the penalty function $f_{R^{JP}}$ (lower panel) as a function of the penalty parameter $\lambda$. The vertical solid line indicates the value of $\lambda$ for which the BIC is minimal.}
\label{fig:KXi-LASSO}
\end{figure}

We emphasize that in the present analysis for determining the minimally required resonance content to describe the data through the LASSO+BIC method, we keep our model as close as possible with that of Ref.~\cite{Jackson:2015dva} apart from the number of resonances considered as described above. For example, the phenomenological contact amplitudes are kept the same expect for the corresponding parameter values 
that are refitted here. Also, the masses and widths of the resonances are kept fixed as in Ref.~\cite{Jackson:2015dva}. Of course, the resonance content depends on whether or not masses and widths are also allowed to vary in the fitting procedure. However, the major motivation here for keeping these parameters fixed is to be able to make a close comparison of the resonance content with the more conventional method of manually determining the resonance content used in Ref.~\cite{Jackson:2015dva}, where these parameters were kept fixed due to the poor quality of the data. Thus, for a meaningful comparison, we perform our analysis under the same constraints.    

Obviously, to determine the resonance content more accurately, we should allow the masses and widths of the resonances to vary as well during the fitting procedure. This, however, may be reserved for a future work when a more accurate and larger database becomes available.

\begin{figure*}[t] \centering
\includegraphics[width=0.48\textwidth,clip=1]{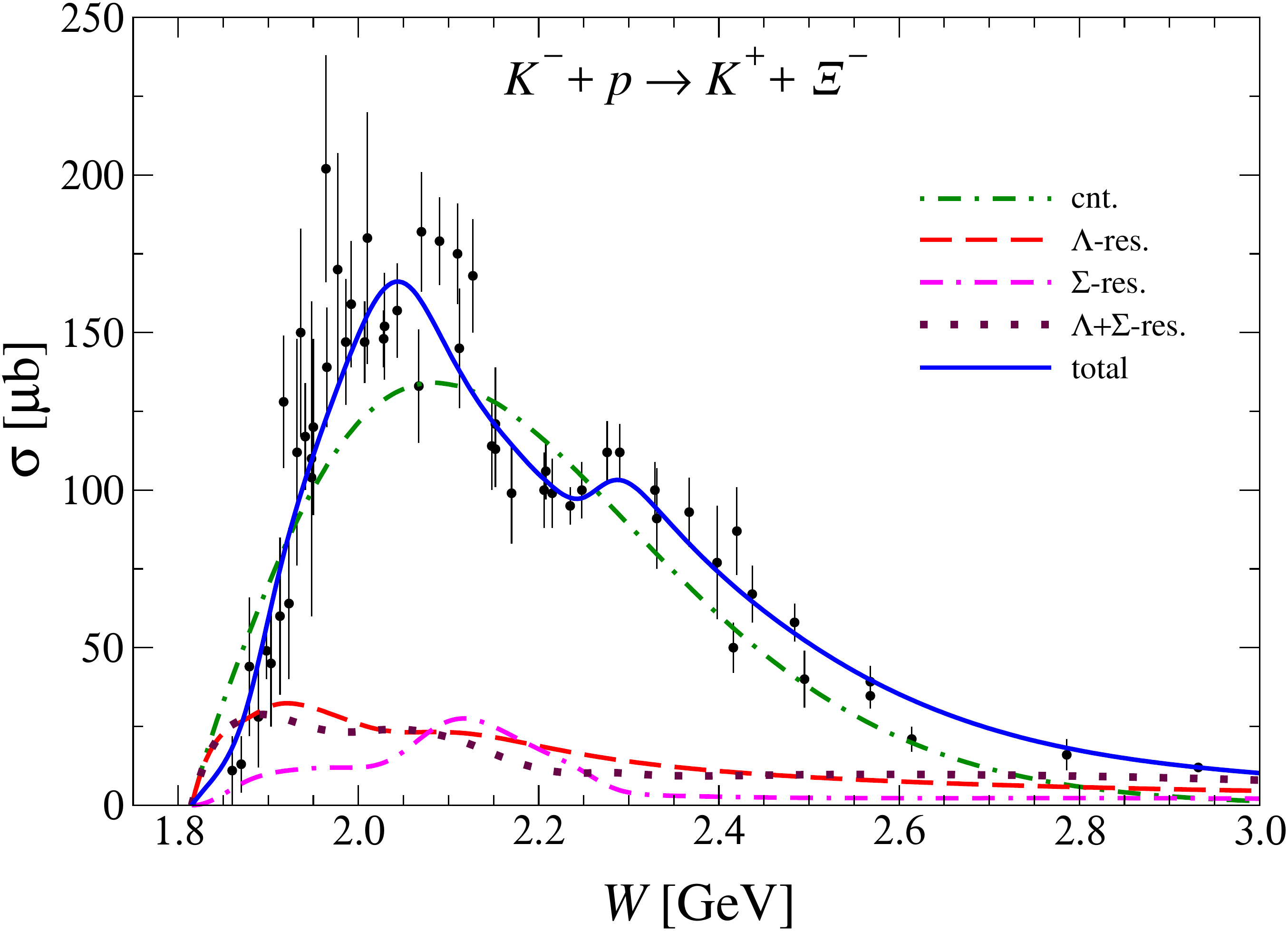} \quad
\includegraphics[width=0.48\textwidth,clip=1]{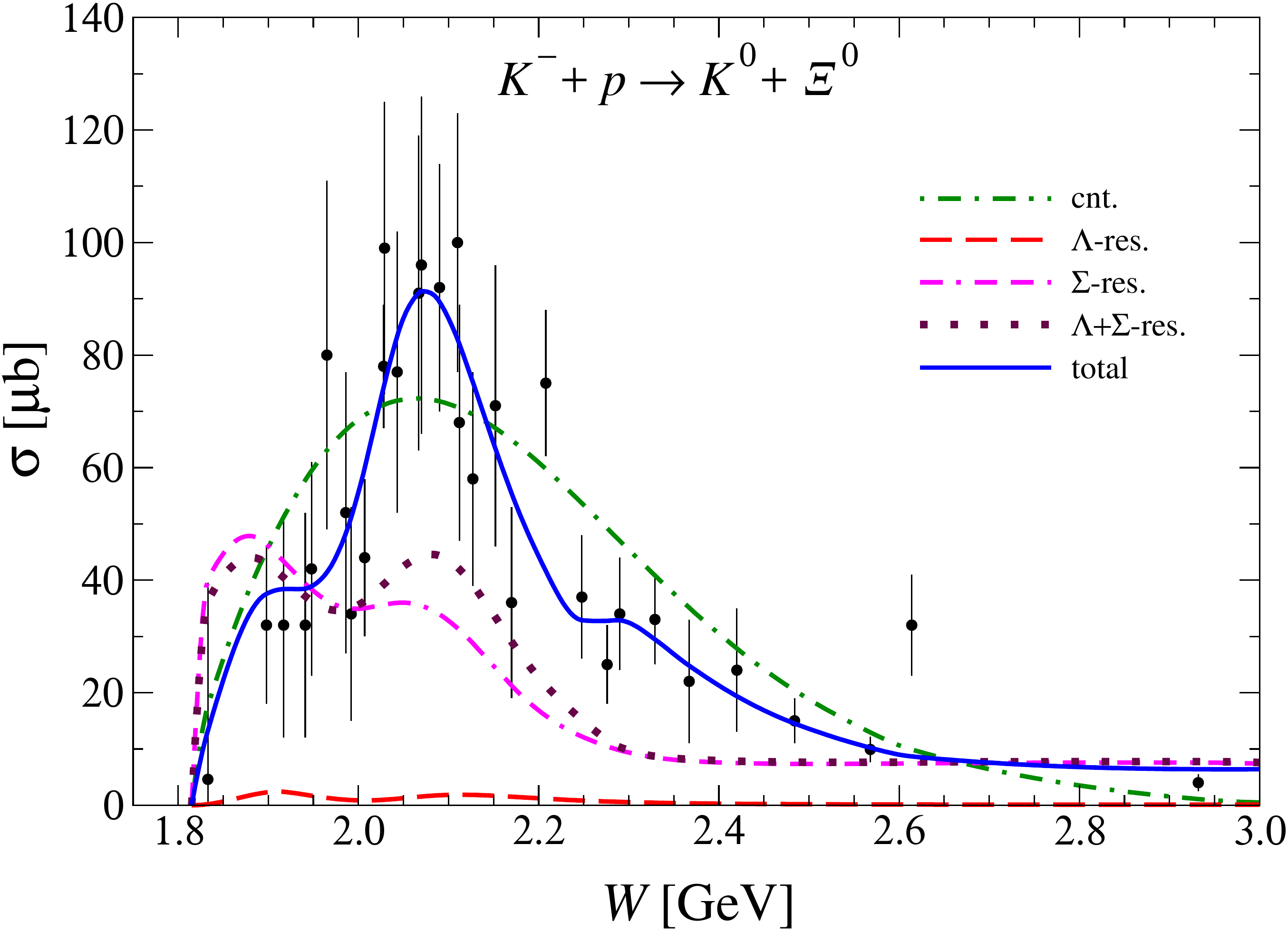}
\qquad
\includegraphics[height=0.31\textwidth,clip=1]{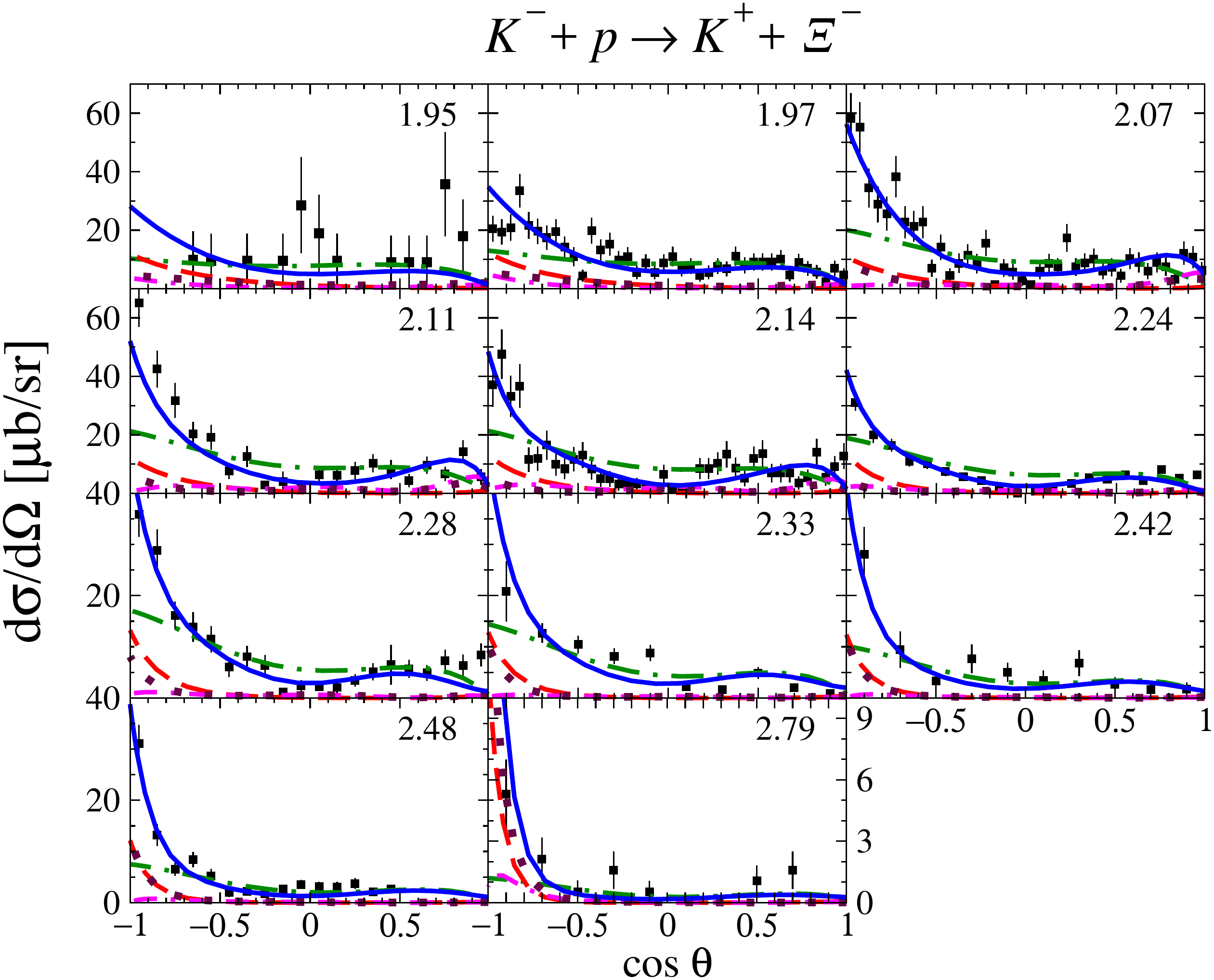}
\includegraphics[height=0.31\textwidth,clip=1]{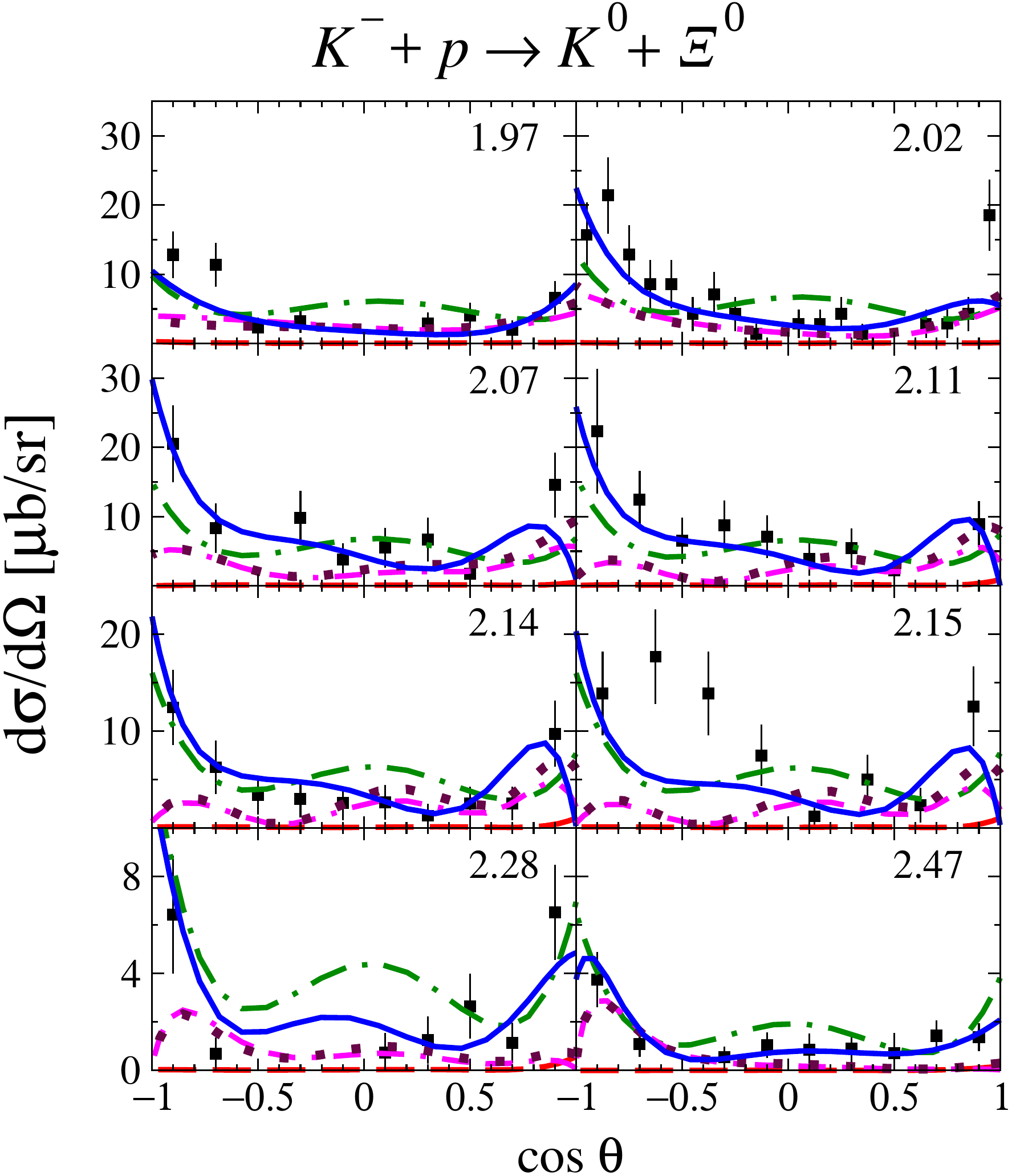}
\includegraphics[height=0.31\textwidth,clip=1]{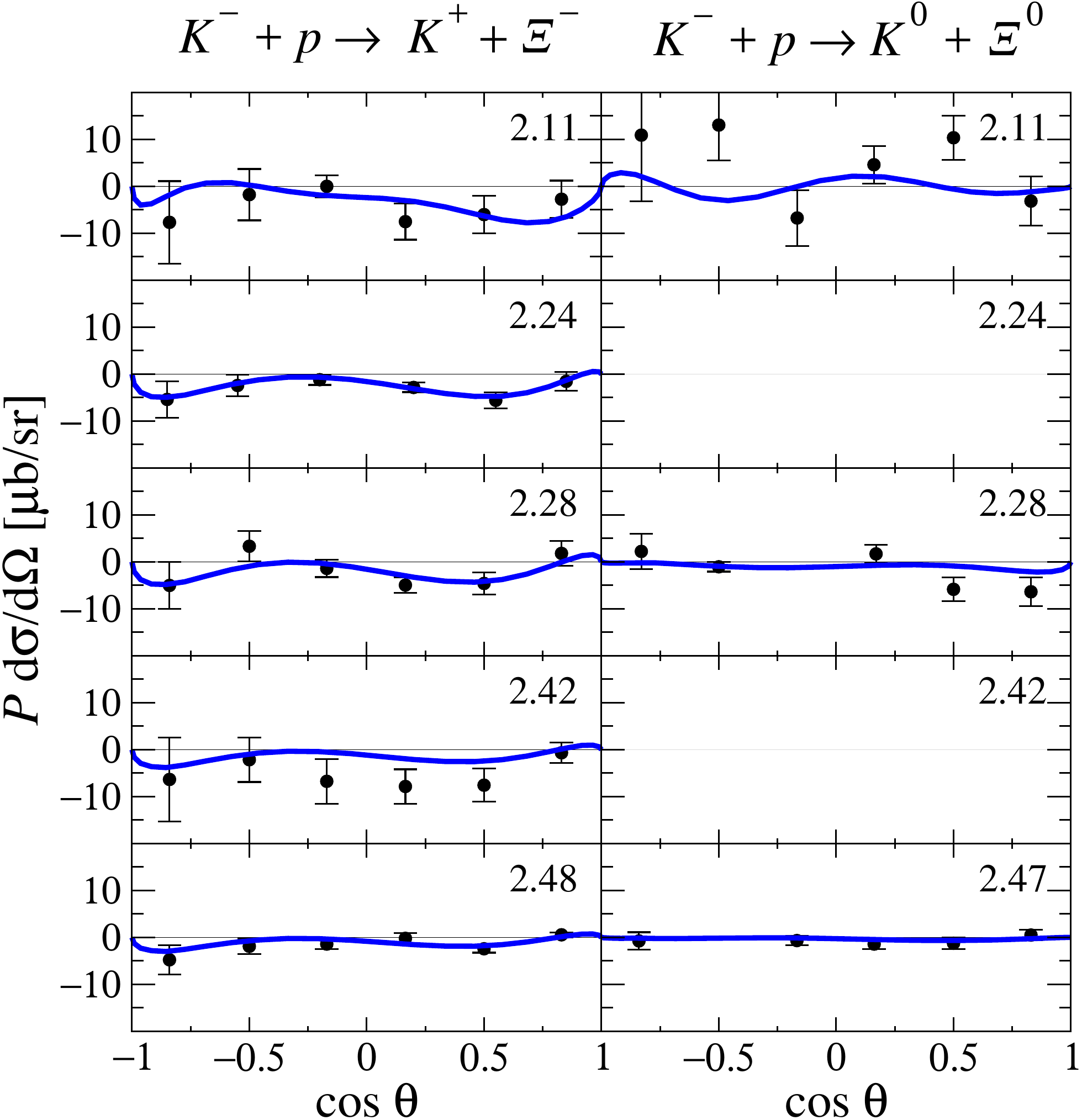}
\caption{
Quality of the model favored by LASSO+BIC for the $K^-p\to K^+\Xi^-$ and $K^-p\to K^0\Xi^0$ reactions compared with the available data~\cite{Carmony:1964zza,Berge:1966zz,Berge:1966zz,London:1966zz,Trippe:1967wat,Burgun:1969ee,Trower:1968zz,Dauber:1969hg,Scheuer:1971zw,DeBellefon:1972gm,Rader:1973ja,Griselin:1975pa,Briefel:1977bp}. The solid blue line represents the result of the full calculation. The red (dashed), magenta (dash-dotted) and brown (dotted) line show the contribution of $\Lambda$, $\Sigma$ and combined $\Lambda/\Sigma$ hyperons, respectively. The green dash-dash-dotted line corresponds to the amplitude with no resonances but contact interactions only.
\label{fig:KXi-txsc}}
\end{figure*}

\subsection{Penalty function for LASSO} 
\label{subsec:KXi-LASSO}

For an above-threshold resonance, the square of its $s$-channel amplitude, when the resonance is on-shell, is proportional to \cite{MON11}
\begin{equation}
\lvert M_{J^{\pm}} \rvert^2 \propto \begin{cases*}
(\varepsilon_N \mp m_R)(\varepsilon_\Xi  \mp m_R)\,,&if $J=\frac12,\frac52$\,,  \\
(\varepsilon_N \pm m_R)(\varepsilon_\Xi  \pm m_R)\,,&if $J=\frac32,\frac72$\,,  
\end{cases*}
\label{eq:KXi-osfactor}
\end{equation}
where $M_{J^P}$ denotes the reaction amplitude involving the intermediate hyperon $R$ with the spin-parity $J^P$;  
$\varepsilon_i \equiv \sqrt{p_i^2 + m_i^2}$, with $p_i$ and $m_i$ denoting the momentum and mass  for $i\in\{N,\Xi\}$, respectively. This proportionality is valid only when the intermediate hyperon lies on its mass shell, and it does not quite apply to the low-mass resonances, which are far off-shell in the present reaction. The above relation shows that the above-threshold unnatural-parity resonances may be suppressed with respect to the natural-parity resonances, unless the corresponding coupling constants are much larger.

In the backward automatic shutoff LASSO method, i.e., starting from a reasonably good local minimum at $\lambda=0$, we minimize the $\chi^2_{T}$ from Eq.~\eqref{totalchi} with the penalty function $P_{J^\pm}(\lambda)=\lambda^2\sum_R \lvert f_{R}\rvert$ with respect to couplings weighted according to Eq.~(\ref{eq:KXi-osfactor}) as
\begin{equation}
f_{R^{J^\pm}}  =  
g_{R^{J^\pm}}\frac{\Gamma_0}{\Gamma_R}
\begin{cases*}
\sqrt{\frac{(\varepsilon_N \mp m_R)(\varepsilon_\Xi  \mp m_R)}{(\varepsilon_N + m_R)(\varepsilon_\Xi  + m_R)}}\,, &  if $J=\frac12, \frac52$ , \\
\sqrt{\frac{(\varepsilon_N \pm m_R)(\varepsilon_\Xi  \pm m_R)}{(\varepsilon_N + m_R)(\varepsilon_\Xi  + m_R)}}\,, &  if $J=\frac32, \frac72$\,, 
\end{cases*}
\label{eq:KXi-PenaltyFunction}
\end{equation}
where $g_{R^{J^\pm}}$ and $\Gamma_R$ stand for the coupling constant and width of the hyperon resonance $R$, respectively. The overall scale normalization is chosen to be $\Gamma_0 = 150$~MeV. 

\subsection{Results}
\label{subsec:KXi-Results}

In this section, we present our results on the resonance content extracted from the available data for the reaction $\bar KN \to K\Xi$ in different isospin channels based on LASSO+BIC. 
The results of LASSO and BIC are collected in Fig.~\ref{fig:KXi-LASSO}. The middle panel shows the result of  BIC with the minimum at $\lambda=\lambda_{\rm opt}\approx 5$. The upper panel displays the $\chi^2_{\rm d.o.f.}$ as a function of the penalty parameter $\lambda$, see Eq.~\eqref{eq:KXi-PenaltyFunction}. The lower panel shows the absolute values of the weighted resonance couplings $f_{R^{J^\pm}}$ as given in Eq.~(\ref{eq:KXi-PenaltyFunction}). According to the BIC, the selected resonances are those whose corresponding weighted couplings  $f_{R^{J^\pm}}$ are above the chosen cutoff of 0.001 at the value of $\lambda$ where the BIC has a  minimum. In Fig.~\ref{fig:KXi-LASSO} we observe at $\lambda=\lambda_{\rm opt}$ a clear distinction between irrelevant resonances ($|f_{R^{JP}}|<10^{-3}$) and relevant ones that all have couplings of size $|f_{R^{JP}}|>10^{-1}$, except for the mentioned $\Sigma(2265)5/2^-$ that shows a small but almost $\lambda$-independent coupling (orange line). Indeed, this resonance produces small but significant bump structures in the data (see Fig.~\ref{fig:KXi-txsc}).
Ten resonances remain out of 21 initial resonances as indicated in Fig.~\ref{fig:KXi-LASSO} (lower panel).

The quality of the results of the model favored by the LASSO+BIC method is illustrated in Fig.~\ref{fig:KXi-txsc}. There, the contributions from those resonances selected by the LASSO+BIC are displayed as red (dashed)  and magenta (double-dash-dotted) curves corresponding to the $\Lambda$ and $\Sigma$ resonances, respectively. The brown (dotted) curves are the total resonance contribution. The green (dash-dotted) curves correspond to the phenomenological contact interaction which accounts effectively for the higher-order (loop) terms in the scattering amplitude~\cite{Jackson:2015dva}. The blue curves correspond to the full total contributions. The overall $\chi^2_{\rm d.o.f.}$ is 2.25.

\begin{table}[t]
\centering
\begin{tabular}{llll}
\hline
resonance switched off~~~~~~
&rating~~~~~~
& $\chi^2_{\rm d.o.f.}$ ~~~~~~
&  $\delta\chi^2 (\%)$  \\
\hline 
none (full result)  &   -     & 2.25     &     -     \\
$\Sigma(2030)7/2^+$ &  ****   & 5.59     &   59.76   \\
$\Sigma(1940)3/2^+$ &       * & 2.49     &   9.60    \\
$\Sigma(2100)7/2^-$ &     *   & 2.46     &   8.36    \\
$\Lambda(2020)7/2^+$&    *    & 2.41     &   6.63    \\
$\Sigma(1840)3/2^+$ &       * & 2.41     &   6.52    \\
$\Lambda(1890)3/2^+$&  ****   & 2.40     &   6.18    \\
$\Sigma(2265)5/2^-$ &   ***   & 2.35     &   4.37    \\
$\Sigma(2070)5/2^+$ &      *  & 2.33     &   3.36    \\
$\Sigma(1915)5/2^+$ &   ****  & 2.29     &   1.69    \\
$\Lambda(2100)7/2^-$&  ****   & 2.26     &   0.48    \\
\hline
\end{tabular}
\caption{
Effects of individual resonances on $\chi^2_{\rm d.o.f.}$ corresponding to Fig.~\ref{fig:KXi-txsc_res}. The third column shows the $\chi^2_{\rm d.o.f.}$ obtained when the corresponding resonance is switched off. 
$\delta\chi^2 \equiv (\chi^2 - \chi^2_{\rm full}) / \chi^2$,\  \ $\chi^2_{\rm full}=2.25$.
\label{tbl:KXi-res}}
\end{table}

\begin{figure*}[t!]\centering
\includegraphics[width=0.49\textwidth,clip=1]{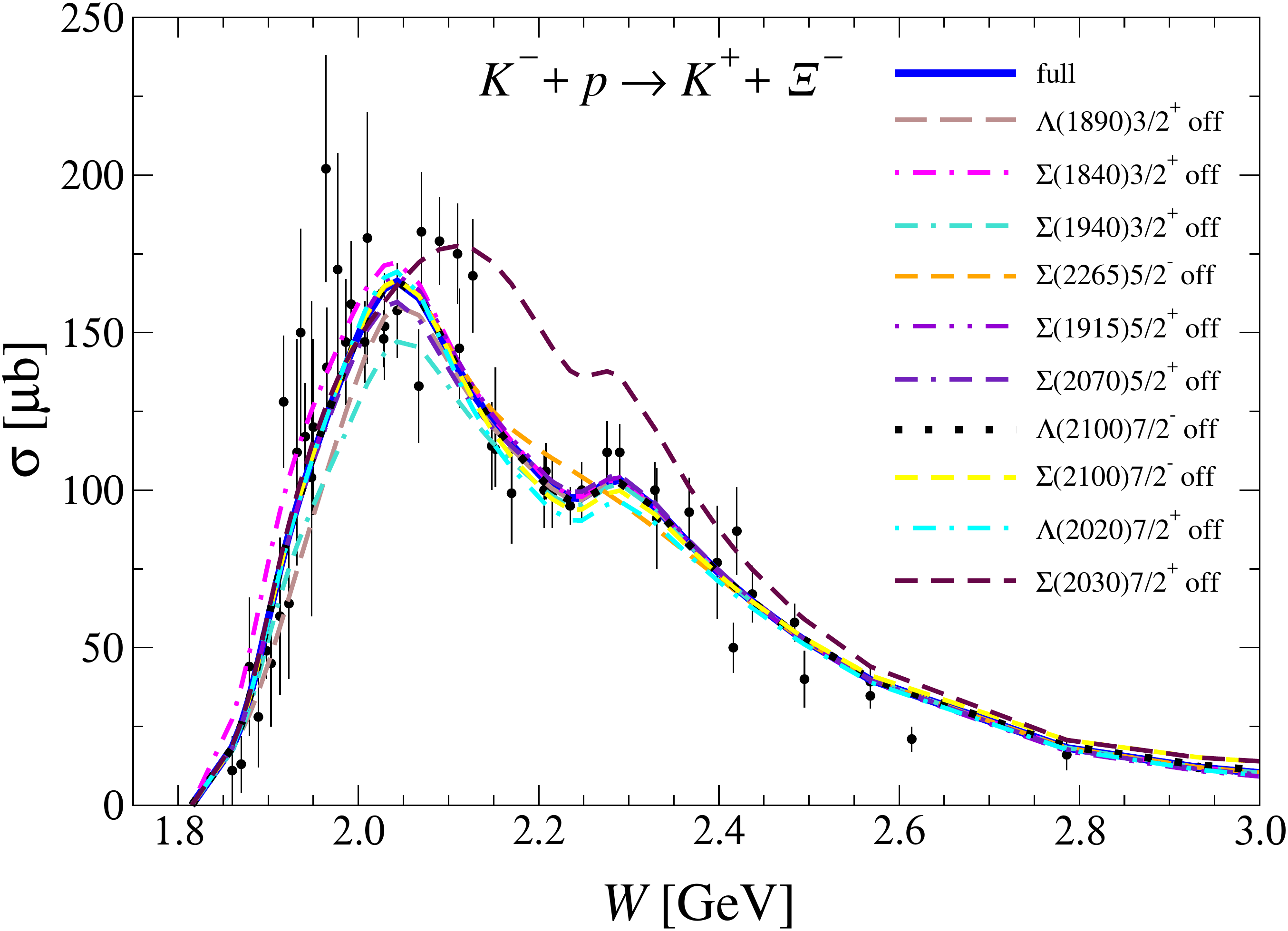} ~
\includegraphics[width=0.49\textwidth,clip=1]{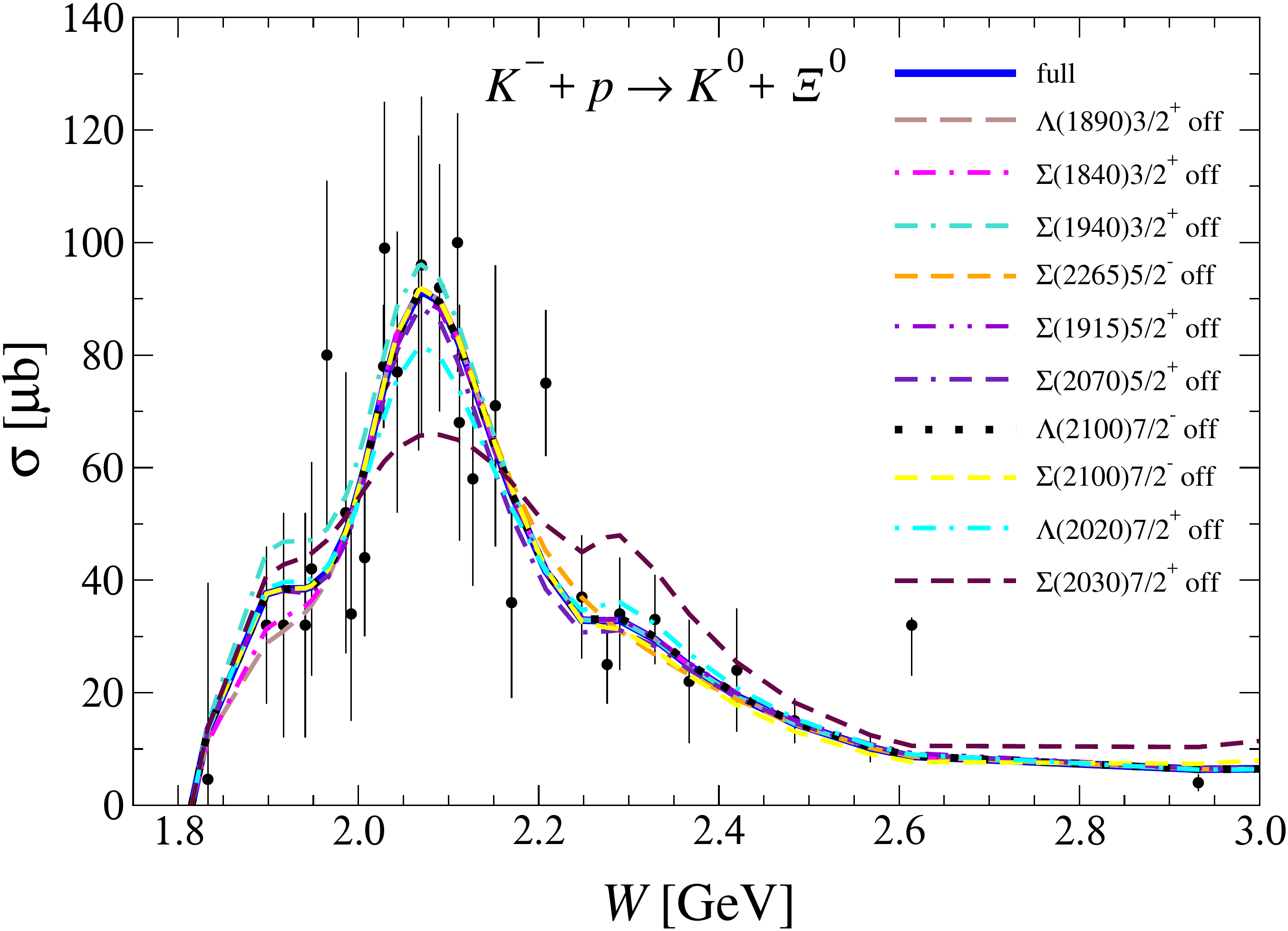}
\caption{
Effects of the individual resonances selected by the LASSO+BIC, as shown in Fig.~\ref{fig:KXi-LASSO} (lower panel), on the total cross sections. $\chi^2_{\rm d.o.f.} \approx 2.25$ with respect to the $3\sigma$-pruned data.
\label{fig:KXi-txsc_res}
}
\end{figure*}

To demonstrate the influence of each resonance (selected by LASSO+BIC), we switch each one off individually, comparing the prediction of the total cross sections as depicted in Fig.~\ref{fig:KXi-txsc_res}. The corresponding numerical changes of the overall $\chi^2_{\rm d.o.f.}$ are collected in Tab.~\ref{tbl:KXi-res}. We see in Fig.~\ref{fig:KXi-txsc_res} that the $\Sigma(2030)$ mostly affects the cross sections in the range of $W \sim 2.0$ to 2.4~GeV. Also, in Tab.~\ref{tbl:KXi-res}, we see that among the ten resonances selected by LASSO+BIC, this resonance affects the overall $\chi^2_{\rm d.o.f.}$ the most (by $\sim 60\%$). It is clearly needed in our model to reproduce the data. Moreover, as pointed out in Ref.~\cite{Jackson:2015dva}, it affects the recoil polarization as well. It should also be mentioned that this resonance helps to reproduce the measured $K^+\Xi^-$ invariant mass distribution in $\gamma p \to K^+ K^+ \Xi^-$ \cite{Man:2011np}, by filling in the valley in the otherwise double-bump structured invariant mass distribution, a feature that is not observed in the data~\cite{Guo:2007dw}. The other resonances have much smaller effects on the total cross sections, as well as on the overall $\chi^2_{\rm d.o.f.}$; the latter  is affected by less than 10\% (cf. Table.~\ref{tbl:KXi-res}).  Five of them  [$\Sigma(1940)3/2^+$, $\Sigma(2100)7/2^-$, $\Lambda(2020)7/2^+$, $\Sigma(1840)3/2^+$, $\Lambda(1890)3/2^+$] affect the $\chi^2_{\rm d.o.f.}$ by about 6\% to 10\%. Here, except for the $\Sigma(1890)3/2^+$ resonance, which has four-star rating, the other four resonances are all one-star resonances. The remaining resonances [$\Sigma(2265)5/2^-$, $\Sigma(2070)5/2^+$, $\Sigma(1915)5/2^+$, $\Lambda(2100)7/2^+$] affect the overall $\chi^2_{\rm d.o.f.}$ by less than 5\%.  In particular, the $\Lambda(2100)7/2^+$ resonance affects the $\chi^2_{\rm d.o.f.}$ by less than 0.5\%. Note that although the $\Sigma(2265)5/2^-$ resonance affects the overall $\chi^2_{\rm d.o.f.}$ by only about 4.4\%, it is very much required to reproduce the small bump structure observed in the total cross section in the $K^- p \to K^+ \Xi^-$ reaction, see Fig.~\ref{fig:KXi-txsc_res} and discussion above. This comparison shows that simple LASSO+BIC resonance selection criterion does not directly translate to the one by examining the total $\chi^2_{\rm d.o.f.}$. Furthermore, the PDG ranking of hyperon resonances is uncorrelated with the LASSO+BIC selection criterion used in this work.

In the analysis of Ref.~\cite{Jackson:2015dva}, the $\Sigma(2030)7/2^+$, $\Sigma(2265)5/2^-$ and $\Lambda(1890)3/2^+$ resonances were identified to be the most relevant ones to reproduce the data. There, only the above-threshold four-star hyperon resonances were considered initially. Then, considering many possible combinations of these resonances, it has been found that the above mentioned three resonances were needed to reproduce the data. In the present analysis, the blindfold search for the above-threshold resonances based on the LASSO+BIC method, also finds these three resonances to be required. However, in addition, the method finds seven more resonances. Among these seven resonances, five are  rated one-star and two are rated four-star. The latter two resonances, $\Sigma(1915)5/2^+$ and $\Lambda(2100)7/2^-$, which have not been found in the analysis of Ref.~\cite{Jackson:2015dva}, however, have only minor influence and affect the overall $\chi^2_{d.o.f.}$ by less than 1.7\% and 0.5\%, respectively.       

To close this section we re-iterate that the result of finding ten relevant resonances depends on (a) the chosen background and (b) whether or not the resonances masses and widths were held constant at their initial values. Choice (a) ensures that results are comparable to Ref.~\cite{Jackson:2015dva} but is, of course, not unique. Restriction (b) is owed to the sparse data for the $K^-p\to K\Xi$ reaction. In general, 
model selection cannot fully address the bias-variance tradeoff that depends on the flexibility of the background parametrization (see also Ref.~\cite{Williams:2017gwf}).

\section{Conclusion}
\label{sec:Conclusions}

Many theory approaches rely on the correct determination of the resonance spectrum from experiment. The Least Absolute Shrinkage and Selection Operator (LASSO) produces, for each penalty $\lambda$, a model with minimal resonance content. As the penalty is convex, the automatized method tests not only resonance by resonance but also combinations thereof --- something that cannot be fully achieved manually. Using synthetic data and criteria from information theory, we have tested forward and backward selection as well as different kinds of penalties. At the given data precision, most variants were able to reproduce the spectrum. Forward selection also provides an efficient way of finding good local minima for this non-linear optimization problem. 

LASSO was then applied to real data of the reaction $K^-p\to K\Xi$. After pruning the data in a self-consistent way to remove outliers, a clear minimum in the Bayesian Information Criterion (BIC) was found, leading to the selection of 10 out of 21 resonances. Remarkably, a minimum in BIC forms even if the $\chi^2$ is not good ($\chi^2_{\rm d.o.f.}\approx 2.3$), i.e., the method seems to be robust. However, while LASSO is a useful tool for model selection, it does not solve the bias-variance problem regarding the parametrization of the background; the challenge persists to construct a parametrization that fulfills as many $S$-matrix properties as possible  to constrain the amplitude.
As an outlook, further testing regarding the impact of systematic uncertainties is advisable as well as the testing of further variants of LASSO versions in connection with stability selection~\cite{Meinshausen2010} to attach probabilities to resonance signals.

\begin{acknowledgments}
The authors thank E. Barut, C. Fern\'andez Ram\'irez and A. Pilloni for discussions.
M.D. acknowledges support by the National Science Foundation (CAREER grant no. PHY-1452055 and PIF grant no. PHY-1415459) and by the U.S. Department of Energy, Office of Science, Office of Nuclear Physics under contract number DE-AC05-06OR23177. M.D. and H.H. acknowledge support by the U.S. Department  Energy, Office of Science, Office of Nuclear Physics under contract number DE-SC0016582.  M.M. is thankful to the German Research Foundation (DFG) for the financial support, under the fellowship MA 7156/1-1, as well as to The George Washington University for the hospitality and inspiring environment.
\end{acknowledgments}

\appendix

\section{Observables}
\label{sec:appA}
For completeness, the observables in terms of partial-wave amplitudes $\tau$ from Eq.~(\ref{tau}) are quoted. The differential cross section $d\sigma/d\Omega$ and polarization $P:=|\vec P_f|$ for an unpolarized target $\vec P_i=0$ are given by
\begin{align}
\frac{d\sigma}{d\Omega}=(|g|^2+|h|^2) \frac{k_f}{k_i}
\,\,\,\,\text{ and }\,\,\,\,
P\frac{d\sigma}{d\Omega}=\frac{k_f}{k_i}(gh^*+g^*h) \,,
\end{align}
where $k_{i/f}$ denotes the magnitude of the initial/final state three-momentum, respectively. The spin-non-flip and spin-flip amplitudes $g_I$ and $h_I$ for the total Isospin $I=0$ and $I=1$ of the reaction $\bar KN\to K\Xi$ can be expressed as an expansion in pertinent partial-wave amplitudes ($\tau_I^{J\pm}$) with respect to the total ($J$) and orbital angular momentum $L$ where the $\pm$ superscript corresponds to $L=J\pm \nicefrac{1}{2}$:
\begin{widetext}
\begin{align}
&g_I=\sum_{J=\nicefrac{1}{2}}^{J_{max}} \frac{(2J+1)}{2 \sqrt{k_f  k_i}} \Big[d^J_{\frac{1}{2} \frac{1}{2}}(\theta)
\cos\left(\frac{\theta}{2}\right)
\left(\tau^{J-}_I+\tau^{J+}_I\right) 
+ d^J_{-\frac{1}{2} \frac{1}{2}}(\theta)
\sin\left(\frac{\theta}{2}\right)
\left(\tau^{J-}_I-\tau^{J+}_I\right)\Big] \ ,\nonumber\\
&h_I=-i\sum_{J=\nicefrac{1}{2}}^{J_{max}} \frac{(2J+1)}{2 \sqrt{k_f  k_i}} \Big[d^J_{\frac{1}{2} \frac{1}{2}}\left(\theta\right)
\sin\left(\frac{\theta}{2}\right)
\left(\tau^{J-}_I+\tau^{J+}_I\right) 
- d^J_{-\frac{1}{2} \frac{1}{2}}(\theta)
\cos\left(\frac{\theta}{2}\right)
\left(\tau^{J-}_I-\tau^{J+}_I\right)\Big]\,.\nonumber
\end{align}
\end{widetext}
The series is truncated at the maximal angular momentum $J_{\rm max}=\frac{5}{2}$ for the analysis of synthetic data (Sec.~\ref{sec:formalism}) and $J_{\rm max}=\frac{7}{2}$  for the real data (Sec.~\ref{sec:real-data}).

\section{Synthetic data}
\label{sec:appB}

\begin{figure*}[thb]
\centering
\includegraphics[width=.49\linewidth]{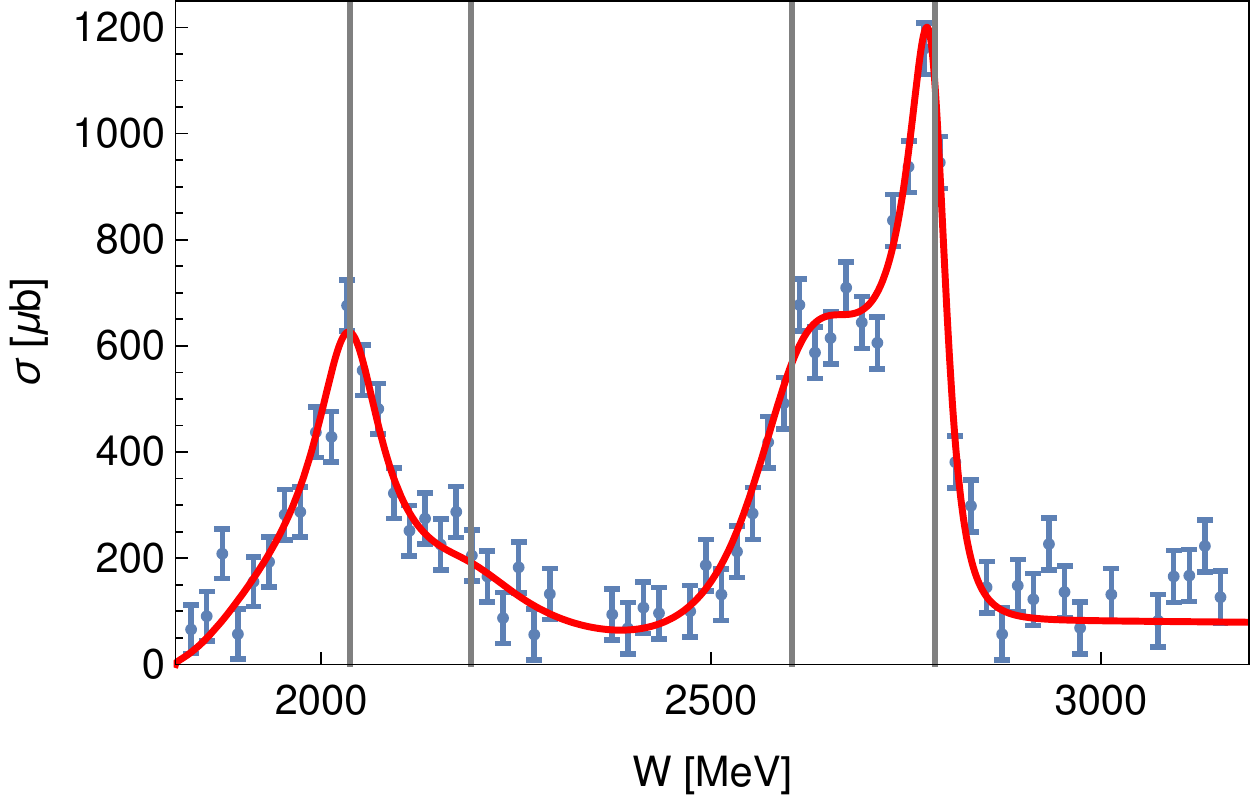}
\includegraphics[width=.49\linewidth]{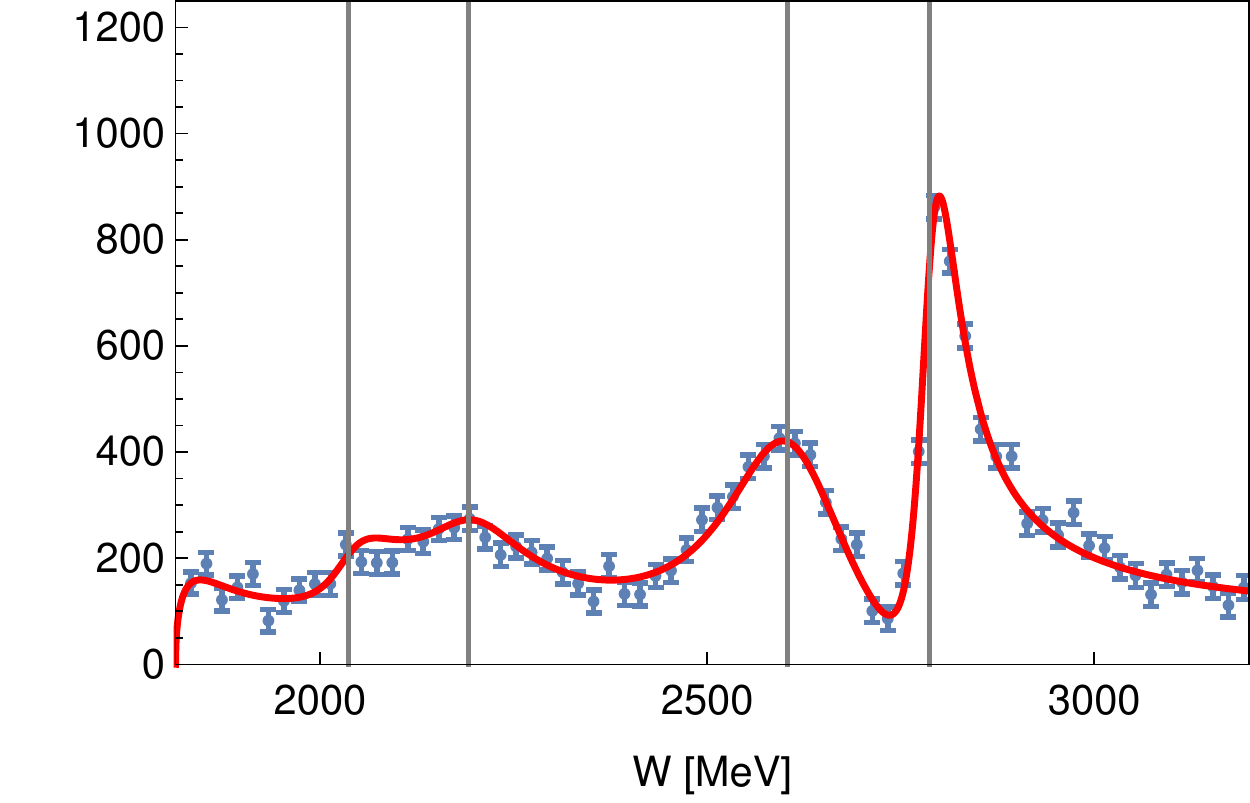}
\caption{Synthetic data (blue dots with error bars) for the total cross sections for the reaction $K^-p\to K^+\Xi^-$ (left) and $K^-p\to K^0\Xi^0$ (right). The generating function is shown in red, while the gray vertical lines depict the resonance masses.}
\label{fig:Fig2}
\end{figure*}

\begin{figure*}
\centering
\includegraphics[width=0.99\linewidth, trim=0.5cm 1cm 1.5cm 1cm]{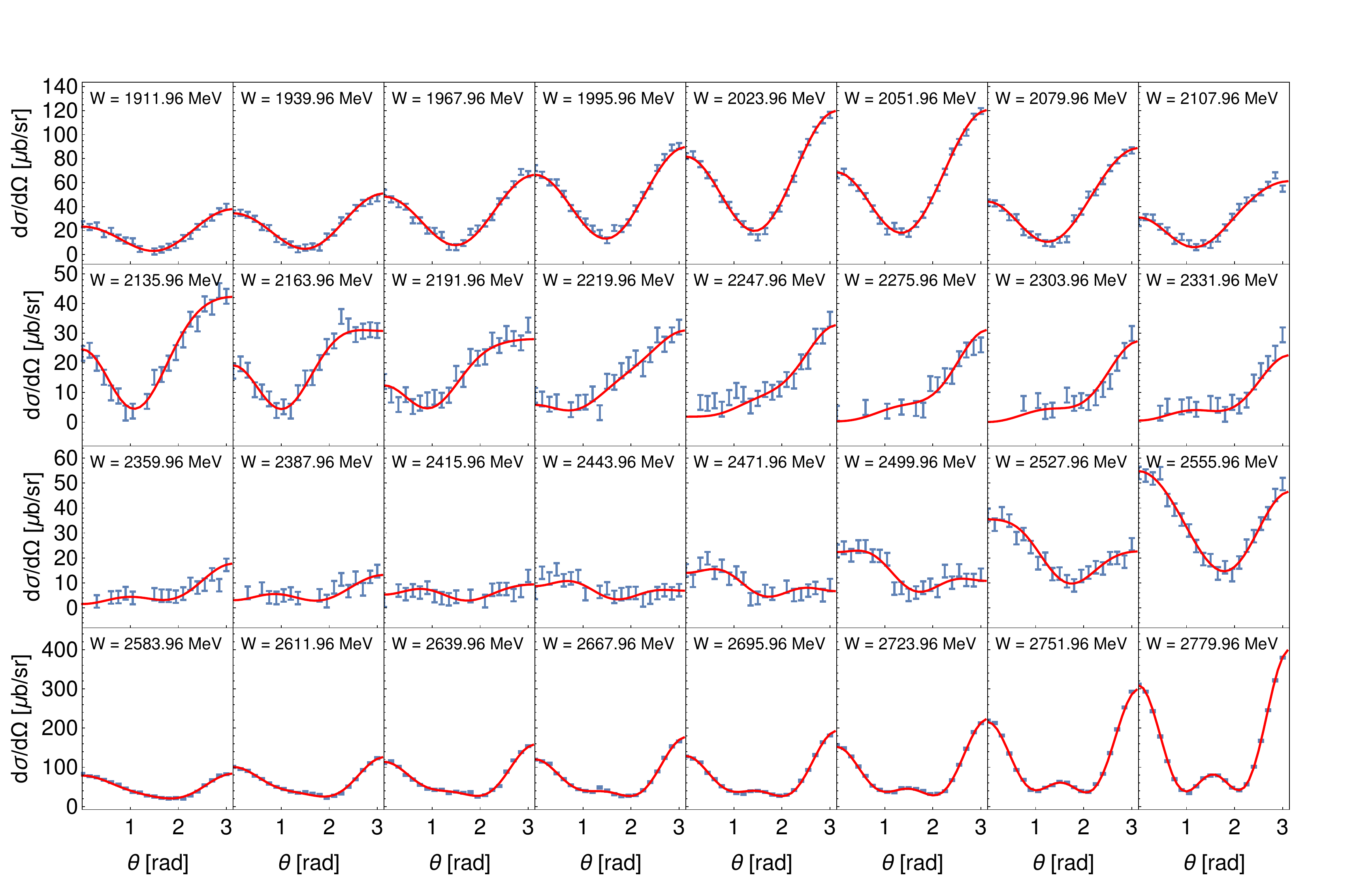}
\includegraphics[width=0.99\linewidth, trim=0.5cm 0.5cm 1.5cm 0cm]{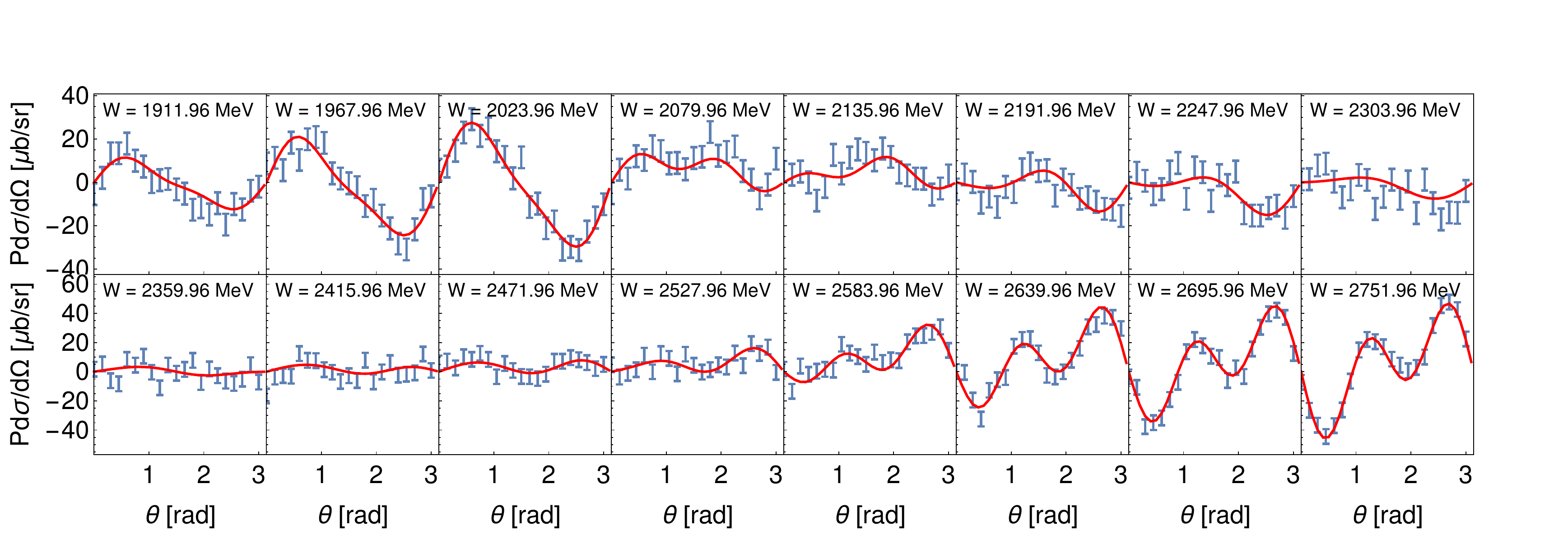}
\caption{Synthetic data for the differential cross sections (top) and polarizations (bottom) for the reaction $K^-p\to K^+\Xi^-$ depicted by blue dots with error bars. The generating function is shown by the red lines.
\label{fig:Fig3}}
\end{figure*}

\begin{figure*}
\centering
\includegraphics[width=0.99\linewidth, trim=0.5cm 1cm 1.5cm 1cm]{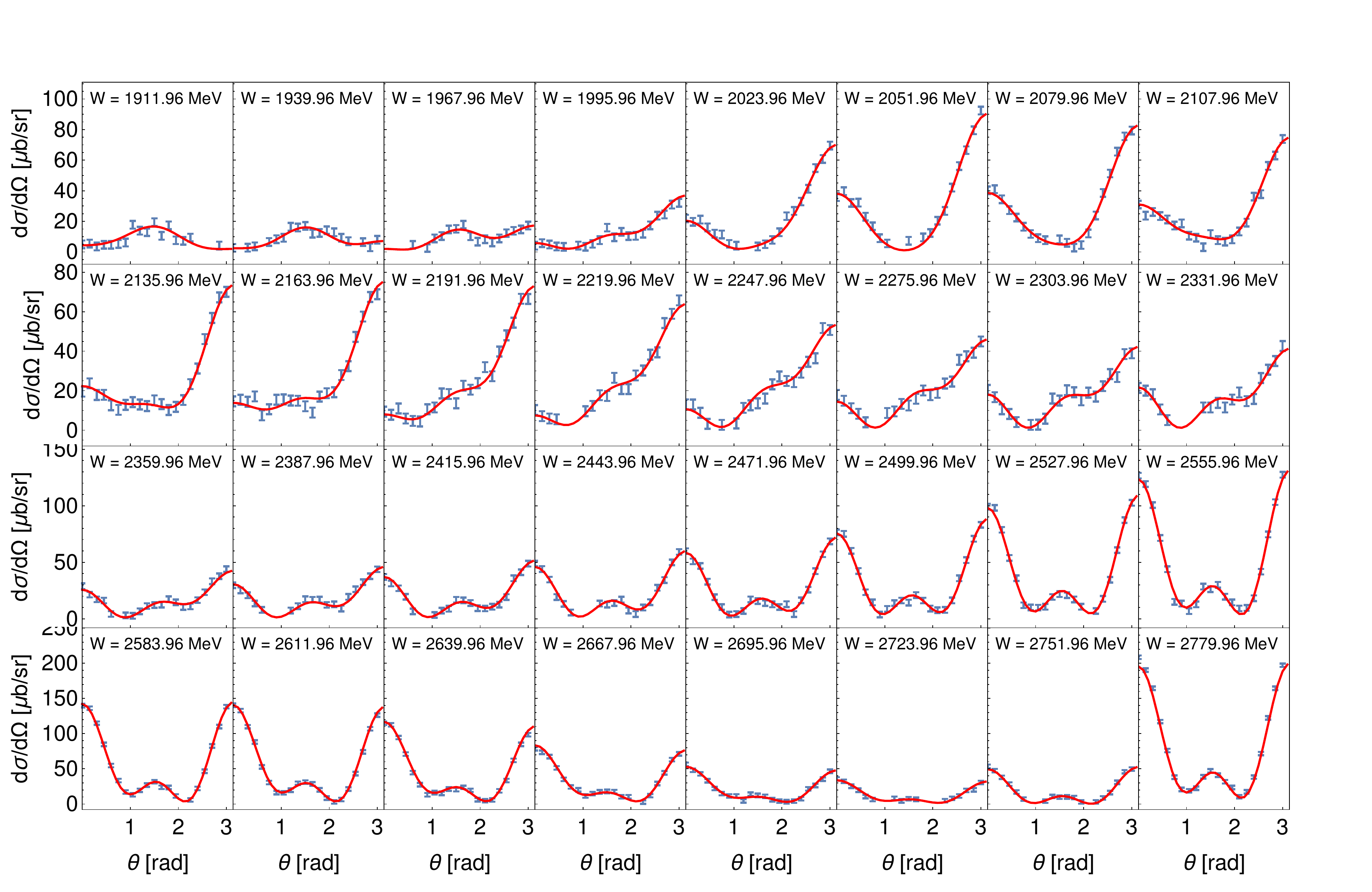}
\includegraphics[width=0.99\linewidth, trim=0.5cm 0.5cm 1.5cm 0cm]{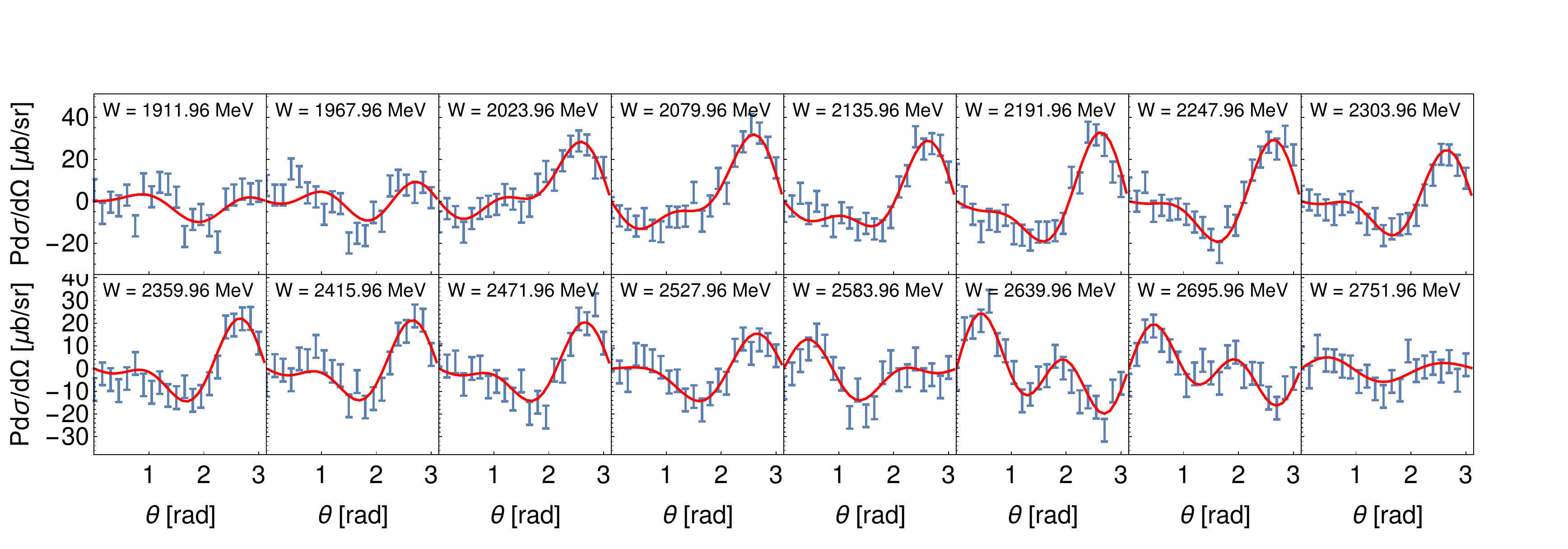}
\caption{Synthetic data for the differential cross sections (top) and polarizations (bottom) for the reaction $K^-p\to K^0\Xi^0$ depicted by blue dots with error bars. The generating function is shown by the red lines.
\label{fig:Fig32}}
\end{figure*}

Figures~\ref{fig:Fig2}--\ref{fig:Fig32} show the synthetic data produced from the partial-waves in Fig.~\ref{fig:PWA-synthetic} as described in the Sec.~\ref{subsec:parametrization}.

\bibliography{stat}
\end{document}